\documentclass[aps,prl,twocolumn,showpacs,superscriptaddress]{revtex4-2}

\usepackage{graphicx}  
\usepackage{hyperref}  
\usepackage{romannum}  
\usepackage[none]{hyphenat}  

\usepackage{soul}
\usepackage{bm}        

\usepackage{amssymb}   

\usepackage[version=4]{mhchem}

\newcommand*\diff{\mathop{}\!\mathrm{d}}

\providecommand{\ehat}{\ensuremath{\hat{\mathbf{e}}}}

\providecommand{\vbPfl}{\ensuremath{ \bar{ \vP }^{(\text{fl})} }}
\providecommand{\vbPfldagger}{\ensuremath{ \bar{ \vP }^{(\text{fl})\dagger} }}
\providecommand{\vbPind}{\ensuremath{ \bar{ \vP }^{(\text{ind})} }}

\newcommand{\veps}{\varepsilon}

\newcommand{\mat}[1]{\boldsymbol{#1}}
\renewcommand{\vec}[1]{\boldsymbol{\mathrm{#1}}}

\newcommand{\bmat}[1]{\begin{bmatrix} #1 \end{bmatrix}}

\providecommand{\vk}{\ensuremath{\vec{k}}}
\renewcommand{\vr}{\ensuremath{\vec{r}}}
\providecommand{\vR}{\ensuremath{\vec{R}}}

\providecommand{\vb}{\ensuremath{\vec{b}}}

\providecommand{\vE}{\ensuremath{\vec{E}}}
\providecommand{\vH}{\ensuremath{\vec{H}}}
\providecommand{\vB}{\ensuremath{\vec{B}}}
\providecommand{\vS}{\ensuremath{\vec{S}}}

\providecommand{\vD}{\ensuremath{\vec{D}}}
\providecommand{\vP}{\ensuremath{\vec{P}}}
\providecommand{\vp}{\ensuremath{\vec{p}}}
\providecommand{\vp}{\ensuremath{\vec{p}}}

\providecommand{\GEE}{\ensuremath{\mat{G}_{EE}}}
\providecommand{\dGEE}{\ensuremath{\text{d}\bar{\mathbb{G}}_{EE}}}
\providecommand{\GHE}{\ensuremath{\mat{G}_{HE}}}

\providecommand{\vbP}{\ensuremath{ \bar{\vec{P}} }}

\providecommand{\vhr}{\ensuremath{ \hat{\vec{r}} }}

\providecommand{\mbT}{\ensuremath{ \bar{\mat{T}} }}
\newcommand{\aver}[1]{\langle #1 \rangle}
\newcommand{\inner}[1]{\langle #1 \rangle}

\renewcommand{\Re}{\mathrm{Re}}

\usepackage{comment}

\begin{document}

\title{Broadband circularly polarized thermal radiation from magnetic Weyl semimetals}


%

\author{Yifan Wang}
\address{School of Electrical and Computer engineering, College of Engineering, Purdue University, West Lafayette, Indiana 47907, USA}

\author{Chinmay Khandekar} \email{ckhandek@stanford.edu}
\address{Department of Physics and Astronomy, Purdue University, West Lafayette, Indiana 47907, USA}
\address{Currently with Department of Electrical Engineering, Stanford University, California 94305, USA}

\author{Xingyu Gao}
\address{Department of Physics and Astronomy, Purdue University, West Lafayette, Indiana 47907, USA}

\author{Tongcang Li}
\address{Department of Physics and Astronomy, Purdue University, West Lafayette, Indiana 47907, USA}
\address{School of Electrical and Computer engineering, College of Engineering, Purdue University, West Lafayette, Indiana 47907, USA}

\author{Dan Jiao}
\address{School of Electrical and Computer engineering, College of Engineering, Purdue University, West Lafayette, Indiana 47907, USA}

\author{Zubin Jacob} \email{zjacob@purdue.edu}
\address{School of Electrical and Computer engineering, College of Engineering, Purdue University, West Lafayette, Indiana 47907, USA}

\date{\today}


\begin{abstract}
We numerically demonstrate that a planar slab made of magnetic Weyl semimetal (a class of topological materials) can emit high-purity circularly polarized (CP) thermal radiation over a broad mid- and long-wave infrared wavelength range for a significant portion of its emission solid angle. This effect fundamentally arises from the strong infrared gyrotropy or nonreciprocity of these materials which primarily depends on the momentum separation between Weyl nodes in the band structure. We clarify the dependence of this effect on the underlying physical parameters and highlight that the spectral bandwidth of CP thermal emission increases with increasing momentum separation between the Weyl nodes. We also demonstrate using recently developed thermal discrete dipole approximation (TDDA) computational method that finite-size bodies of magnetic Weyl semimetals can emit spectrally broadband CP thermal light, albeit over smaller portion of the emission solid angle compared to the planar slabs. Our work identifies unique fundamental and technological prospects of magnetic Weyl semimetals for engineering thermal radiation and designing efficient CP light sources.
\end{abstract}

 \maketitle

\section{Introduction}

Circularly polarized (CP) light is obtained by utilizing either linear to circular polarization conversion device (quarter wave plate) or electroluminescence, photoluminescence, incandescence (thermal radiation) of chiral geometric or material systems
\cite{Nishizawa2017,Asshoff2011,DiNuzzo2017,Zhao2016,Zhang2014,Kumar2015,Sanchez-Carnerero2015,Konishi2011,Maksimov2014,Wadsworth2011,Shitrit2013}
. While a majority of research on CP light sources covers the microwave and near-infrared to visible wavelength range because of many practical applications, mid- and long-wave infrared wavelength range remains relatively less explored. Since thermal radiation proportional to the blackbody distribution peaks in this wavelength range ($3-20\mu$m) near and above room temperature, polarized thermal light sources can be useful for practical applications like infrared chiral spectroscopy \cite{infrared_chiral_spectroscopy, CP_spectroscopy,ranjbar2009circular,basiri2019nature} and thermal imaging polarimetry under low visibility conditions \cite{Hildebrand2000,Gade2014,snik2014overview}. In this work, we reveal that recently discovered magnetic Weyl semimetals, which are strong gyroelectric or magneto-optic materials, can emit strong circularly polarized thermal radiation which is broadband in the mid- and long-wave infrared wavelength range.

Although linearly polarized thermal radiation has been demonstrated using patterned or microstructured geometries\cite{Bimonte2009,Marquier2008}, realizing and detecting circular polarization of thermal emission is more difficult and has been demonstrated in very few experiments using either a chiral absorber metasurface~\cite{Hasman_group_2010} or a polarization conversion metasurface~\cite{Wadsworth2011}. Apart from these experiments and many theoretical proposals relying on nanophotonic design of reciprocal chiral absorbers\cite{Yin2013,Dyakov2018,Lee2007}, new approaches based on nonequilibrium coupled antennas~\cite{chinmay_dipolar2019}  and nonreciprocal media~\cite{Maghrebi_2019,Narimanov2019,Chinmay2020} have been recently explored for CP thermal emission. However, in all of these works so far, either the purity of circular polarization is low or high-purity circular polarization is achieved over configuration-dependent narrow spectral and angular bandwidths. Here, purity can be quantified as the ratio $(I_{\mathrm{(LCP)}}-I_{\mathrm{(RCP)}})/(I_{\mathrm{(LCP)}}+I_{\mathrm{(RCP)}})$ where LCP/RCP denotes left/right circularly polarized light and $I$ denotes the intensity. In Ref.\cite{Wadsworth2011}, $28\%$ purity CP thermal emission in normal direction from a metasurface is realized over $8-12\mu$m range while in other works~\cite{Hasman_group_2010,Dyakov2018,chinmay_dipolar2019,Yin2013}, greater than $50\%$ purity CP thermal emission is reported but over a small wavelength band of $1-2\mu$m and only along specific emission directions. In this work, we show that a rather simple geometry of a planar slab made of a representative magnetic Weyl semimetal~\cite{BoZhao2020}, without any patterning or application of external magnetic field, can lead to greater than $50\%$ purity CP thermal emission over a bandwidth $6-14\mu$m. for around $75\%$ of its emission solid angle.

\begin{figure}[t!]
\centering\includegraphics[width=\linewidth]{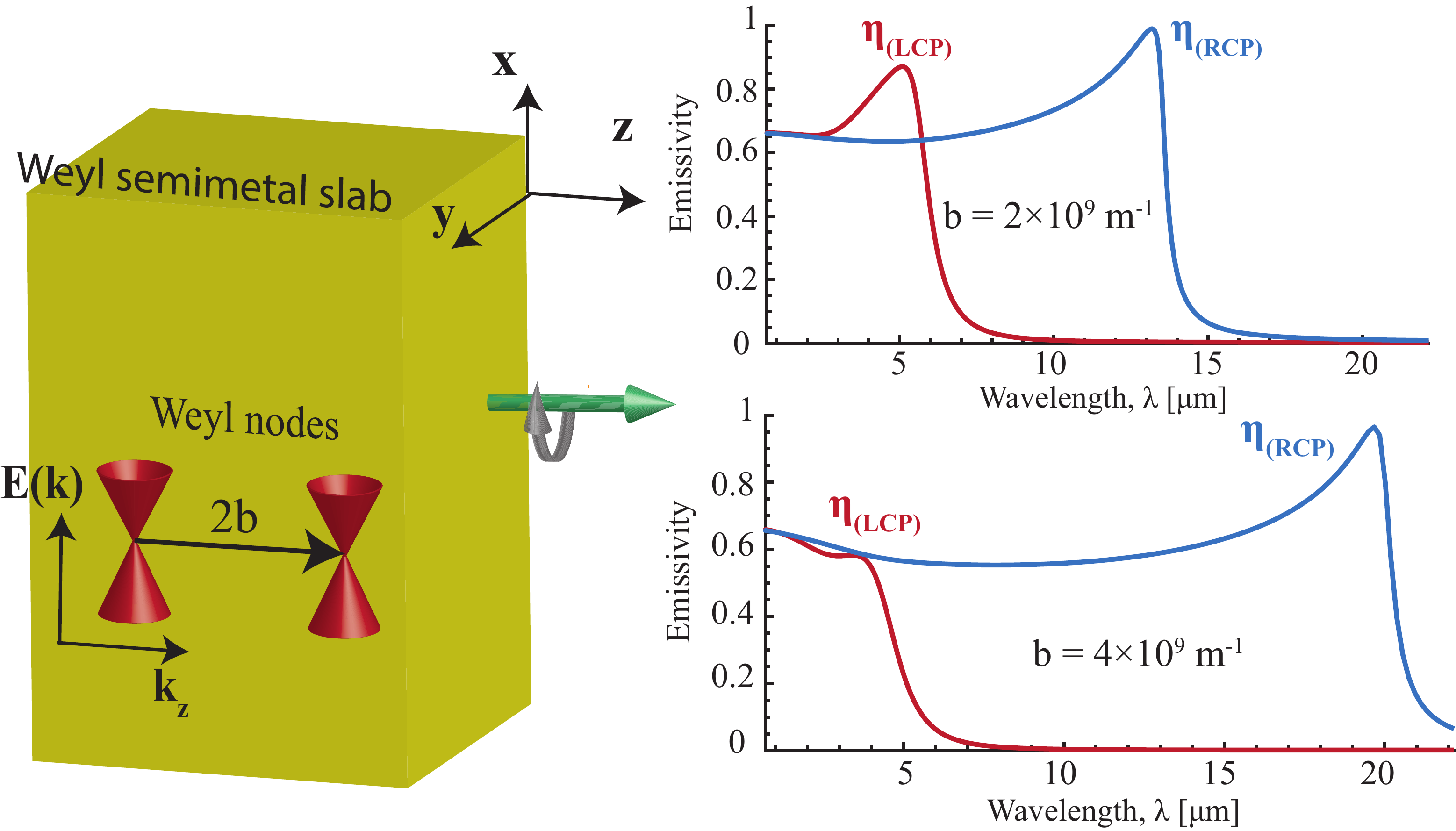}
\caption{We consider a semi-infinite half-space of magnetic Weyl semimetal with infinite transverse extent in the $\mathbf{xy}$ plane. For the Weyl semimetal, we consider the momentum separation of magnitude $2b$ along $\hat{k}_z$ direction as indicated. We show that thermal radiation into the half-space above the surface is strongly circularly polarized due to strong gyrotropy which depends on the momentum separation parameter $b$. The right figures depict the spectra of RCP and LCP light emissivities in the normal direction for a representative example of a Weyl semimetal~\cite{BoZhao2020}, indicating the broadband nature of circular polarization of thermal emission. The figures plotted for two different values of $b$ illustrate that the bandwidth of CP emission increases with increasing momentum separation between Weyl nodes.}
\label{fig1}
\end{figure}

Weyl semimetals are topological quantum materials whose nontrivial bulk band topology manifests as unique physical effects such as the topologically protected Fermi arcs, the chiral anomaly in the bulk, and large negative magnetoresistance \cite{Jia2016, Huang2015}. The topological properties of Weyl semimetals exist only in materials which lack either time-reversal symmetry (magnetic Weyl semimetals) or inversion symmetry (noncentrosymmetric Weyl semimetals) or both.  In this work, we only focus on magnetic Weyl semimetals because of their giant nonreciprocity and henceforth, we abbreviate them as MWS for brevity. Theoretically, there are many candidate MWS materials~\cite{Wan2011,Sushkov2015,Xu2011,kubler2012berry,chang2016room}, and some of them such as \ce{Co3Sn2S2}\cite{Liu_Co3Sn2S2_exp, Morali_Co3Sn2S2_exp}, \ce{Co2MnGa}\cite{Belopolski_Co2MnGa_exp}, \ce{Co2MnAl}~\cite{li2020giant} are confirmed as MWS in recent experiments. While non-centrosymmetric Weyl semimetals are promising for large optical nonlinearity \cite{Morimoto2016}, large photovoltaic\cite{WSM_photovoltaic} and photogalvanic \cite{WSM_photogalvanic} effects, MWS are promising because of their giant nonreciprocity at infrared wavelengths which is unmatched by the existing material options of doped semiconductors in external magnetic field and conventional magnetic materials~\cite{Kotov2016,Kotov2018}. Because of this advantage which stems from the large intrinsic anomalous Hall effect, the nonreciprocity of MWS has been recently exploited for designing ultrasmall footprint optical isolators~\cite{Optical_Isolators} and for optimizing thermal energy harvesting by nonreciprocal violation of directional Kirchhoff’s law of thermal radiation~\cite{Green_Martin_PV, Tsurimaki_Chen_Gang_2020,BoZhao2020}. Our work highlights yet another unique prospect of utilizing giant nonreciprocity of MWS for realizing broadband CP infrared radiation. For planar slabs, we demonstrate that the bandwidth of CP thermal light is strongly correlated with the separation of Weyl nodes in the momentum space. Figure~\ref{fig1} displays the planar geometry considered in this work. Using a representative Weyl semimetal analyzed in Ref.\cite{BoZhao2020}, the figure plots the spectra of RCP and LCP emissivities ($\eta_{\mathrm{(RCP/LCP)}} \in [0,1]$) along the normal emission direction for momentum separations given by $b=2$nm$^{-1}$ (top) and $b=4$nm$^{-1}$ (bottom). Larger the momentum separation between the Weyl nodes, larger is the bandwidth of emitted CP thermal light. We elucidate the dependence of this result on various physical parameters describing the MWS such as Fermi energy, number of Weyl nodes, background permittivity etc., and point out the favorable parameters for achieving larger bandwidths.

We also demonstrate the circular polarization of thermal emission from finite subwavelength as well as wavelength-size bodies made of MWS. We use the thermal discrete dipole approximation (TDDA) method suitable for thermal radiation calculations from finite nonreciprocal bodies \cite{TDDA_2017}, and extend it to analyze the spin or circular polarization of thermal radiation from cube and rod geometries. Our computational simulations and analysis show that strong circular polarization (spin) of thermal radiation is not limited to an extended planar slab but can also be observed with finite-size objects made of MWS.

We note that chiral absorption of Weyl semimetals has been noted recently~\cite{Chiral_anomaly-optical_absorption, Tunable-circular-dichroism_WSM}. However, the implications for CP thermal radiation and its broadband nature have not been reported till date. A chiral absorber can emit partially CP thermal radiation based on a version of Kirchhoff’s law which equates emissivity with absorptivity separately for RCP and LCP light. However, this photon-spin-resolved Kirchhoff’s law is applicable only for reciprocal media~\cite{resnick1999polarized,zhang2020validity}. It does not hold true for nonreciprocal media such as MWS considered here. Interestingly, nonreciprocal media follow different modified forms of spin-resolved Kirchhoff's laws~\cite{Chinmay2020}. The theoretical techniques and computational tools for analyzing the spin or circular polarization of thermal radiation from nonreciprocal bodies are only recently developed in the field of thermal radiation~\cite{Ott2018,Chinmay2020,Chinmay2019}. All prior inquiries concerning the topic of thermal spin photonics analyzed reciprocal media for which analytic derivations and computational simulations are simplified by invoking the principle of electromagnetic reciprocity. That is no longer the case for nonreciprocal media.

The manuscript is organized as the following. We provide the main results in section (\ref{results}) followed by the conclusion (\ref{conclusion}) and the appendix (\ref{appendix}). In section (\ref{results}), we provide the description of optical response or permittivity of MWS (\ref{MWS}) followed by the detailed analysis of spin or circular polarization of thermal emission from extended planar slabs (\ref{EPS}) and finite-size objects (\ref{finite}).

\section{Results}
\label{results}

\subsection{Magnetic Weyl semimetals}
\label{MWS}

There has been a significant body of research in the past decade devoted to the fundamental understanding of Weyl semimetals. They are now theoretically well-understood~\cite{Weyl_RMP}. The band structure of Weyl semimetals contains an even number of nondegenerate band-touching points called as Weyl nodes which are topologically stable. These Weyl nodes appear in pairs of two nodes of opposite chirality separated by a vector $2\vb$ in momentum space and by a scalar $2\hbar b_0$ in energy space (chiral chemical potential). The electromagnetic response of Weyl semimetals is described using the formalism of axion electrodynamics where these parameters appear in the form of axion term in the Lagrangian density. The axion term eventually leads to the following modified constitutive relation for the displacement field in the frequency domain~\cite{BoZhao2020, Hofmann_2016_SPP}:
\begin{equation}
\vD = \veps_d(\omega) \vE + \frac{ie^2}{4 \pi^2 \hbar \omega} (-2b_0\vB + 2\vb \times \vE)
\end{equation}
where $\veps_d(\omega)$ is the frequency-dependent permittivity of the corresponding Dirac semimetal, which has $b_0 = \vb = 0$.
In this work, we consider only materials where Weyl nodes have the same energy such that $b_0=0$. Throughout the work, we choose $\vb = b \hat{\vk}_z$ to be along the $\hat{\vk}_z$ direction of the coordinate system but also mention the implications of other directionalities. Based on these considerations, the permittivity tensor of the MWS is:
\begin{equation}
\varepsilon =
 \bmat{\varepsilon_d & i\varepsilon_a & 0 \\
 -i\varepsilon_a & \varepsilon_d & 0 \\
 0 & 0 & \varepsilon_d \\}
\end{equation}
where
\begin{equation}
\varepsilon_a = \frac{b e^2}{2 \pi^2 \hbar \omega} .
\end{equation}
The presence of off-diagonal anti-symmetric entries in the permittivity tensor leads to electromagnetic nonreciprocity~\cite{Caloz2018}. Such a material is called as gyrotropic medium with gyrotropy axis along $\mathbf{z}$ direction leading to nonzero $\veps_{xy}, \veps_{yx}$ terms. The diagonal part of the permittivity  $\varepsilon_d$ is \cite{BoZhao2020,Kotov2016,Kotov2018}:
\begin{equation}
\varepsilon_d(\omega) = \varepsilon_b + i\frac{\sigma(\omega)}{\omega} .
\end{equation}
Here $\varepsilon_b$ is the constant background permittivity, and $\sigma$ is the bulk conductivity, whose dispersion is given by
\begin{align}
\label{conductivity_eq}
\sigma = &\frac{r_s g}{6} \Omega G(\Omega/2) + i \frac{r_s g}{6\pi} \{
    \frac{4}{\Omega} [
        1+\frac{\pi^2}{3} (\frac{k_B T}{E_F(T)})^2
    ]
    + \nonumber \\
    &8 \Omega \int_0^{\xi_c}
        \frac{G(\xi) - G(\Omega/2)}{\Omega^2-4\xi^2} \xi\ \diff\xi
    \},
\end{align}
$\Omega = \hbar(\omega + i \tau^{-1}) / E_F(T)$ is a complex frequency normalized by Fermi energy $E_F(T)$, $\tau^{-1}$ is the scattering rate corresponding to Drude damping, $G(E) = n(-E) - n(E)$, where $n(E) = (e^{(E-E_F)/(k_BT)}+1)^{-1}$ is the Fermi distribution function, $r_s = e^2/(4\pi\varepsilon_0\hbar\nu_F)$ is the effective fine structure constant, $\nu_F$ is Fermi velocity, $g$ is the number of Weyl nodes, $\xi_c = E_c / E_F$ is the cutoff energy, beyond which the band dispersion is no longer linear \cite{Kotov2016}. In the following, we first use a representative example with parameters similar to Ref.\cite{BoZhao2020} as $\varepsilon_b = 6.2$, $\tau=1000$ fs, $\xi_c = 3$, $b=2\times 10^9$ m$^{-1}$, $\nu_F = 0.83\times 10^5$ m/s and Fermi energy $E_F(T)=0.15$ eV at $T=300$ K. Then we vary these parameters within the range of reported values in the literature and analyze the circular polarization of thermal emission.

\subsection{Extended Planar Slabs}
\label{EPS}

\begin{figure*}[ht!]
\centering\includegraphics[width=0.88\linewidth]{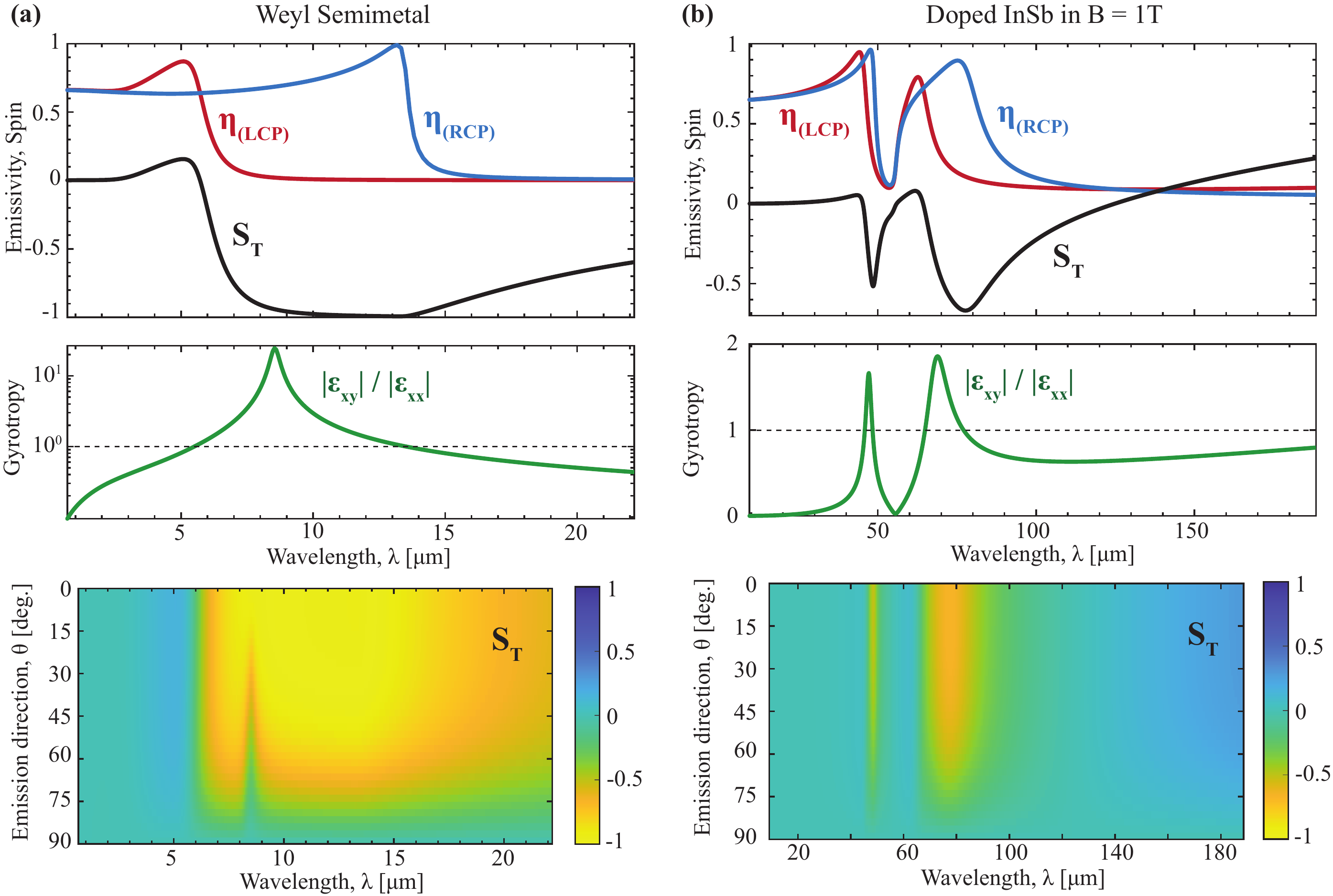}
\caption{The figure compares circularly polarized thermal radiation from thick planar slabs (infinite in $\mathbf{xy}$ plane) of two nonreciprocal media namely (a) magnetic Weyl semimetals with $\vb \parallel \hat{\vk}_z$ and (b) a doped Indium Antimonide (InSb) slab in magnetic field $B=1$T along $\mathbf{z}$. For both (a) and (b), the top figure shows the spectrum of emissivities of RCP (blue) and LCP (red) light and thermal photon spin $S_T$ (black), defined by Eq.\ref{S_T_planar}, for thermal light emitted normal to the surface. The middle figure plots the gyrotropy or equivalently, the strength of nonreciprocity in the same wavelength range. The bottom figure displays the dependence of $S_T$ on the wavelength as well as the emission direction characterized by the angle $\theta$ made by the emission direction with the normal to the surface.}
\label{fig2}
\end{figure*}

We consider a semi-infinite half space (thick planar slab) of representative MWS (parameters noted above) at temperature $T=300$K emitting thermal radiation into the vacuum half space (environment) at $T_0=0$K. The infinite transverse extent lies in the $\mathbf{xy}$ plane with normal to the surface along $\mathbf{z}$ direction. We analyze thermal emission in the direction characterized by the angles ($\theta,\phi$) where $\theta$ is the angle with $\mathbf{z}$ axis and $\phi$ is the angle made by the in-plane component of the emission-direction vector with $\mathbf{x}$ axis. The thermal radiation power emitted per unit surface area $\mathrm{d} A$ of the slab in the direction $(\theta,\phi)$ per unit solid angle $\mathrm{d} \Omega$ per unit wavelength $\mathrm{d} \lambda$ for a given polarization state $\ehat$ is given by,
\begin{equation}
P_{\text{rad}}(\theta, \phi, \lambda, \ehat) = \eta(\theta, \phi, \lambda, \ehat)
\frac{I_{\text{bb}}(\lambda, T)}{2}
\cos(\theta) \diff \lambda \diff \Omega \diff A ,
\label{Prad_planar}
\end{equation}
where $\eta$ is the dimensionless emissivity $\in [0,1]$,  $I_{\text{bb}}(\lambda, T) =
\frac{2hc^{2}}{\lambda ^{5}}
\frac{1}{ e^{hc/(\lambda k_{\mathrm{B}} T)}-1} $ is the blackbody radiance at temperature $T$. It is divided by $2$ in Eq.\ref{Prad_planar} to account for two orthogonal polarization states separately. Since we are interested in the circular polarization of light, the polarization-dependent emissivity is calculated in the eigenbasis of right circularly polarized (RCP) and left circularly polarized (LCP) photons~\cite{Chinmay2020}. Based on the definitions of the polarization eigenstates in vacuum and the definition of the spin angular momentum density \cite{Bliokh2017,Barnett2016,yang2020quantum}, a single RCP photon carries $-\hbar$ angular momentum and a single LCP photon carries $+\hbar$ angular momentum along the propagation direction (see appendix for convention details). We then define thermal photon spin denoted as $S_T$ for the extended planar slabs as:
\begin{equation}
\label{S_T_planar}
S_T(\theta, \phi, \lambda) = \frac{\eta_{(\text{LCP})} - \eta_{(\text{RCP})}}
{\eta_{(\text{LCP})} + \eta_{(\text{RCP})}}
\end{equation}
where $\eta_{\mathrm{(LCP/RCP)}}$ denotes the emissivity of LCP/RCP light. Its dependence on $(\theta,\phi,\lambda)$ in the above equation is omitted for brevity. This quantity is similar to the circular dichroism parameter considered in  circular dichroism spectroscopy~\cite{Ranjbar2009,gottarelli2008use} or the well-known Stokes $S_3$ parameter \cite{jackson1999}. It is also same as the purity of circular polarization described in the introduction. $S_T$ furthermore quantifies the net longitudinal spin angular momentum per emitted thermal photon. It can be measured experimentally using quarter wave plates and detectors oriented in the given emission direction as done in previous experiments~\cite{Wadsworth2011}. The spin-resolved emissivities $\eta_{(\text{LCP/RCP})}$ in the above expressions for nonreciprocal planar media have been derived previously in Ref.\cite{Chinmay2020}, and are separately discussed in the appendix (Section \ref{appendix}) in the context of this work.

In Fig. \ref{fig2}a, the top figure shows the spectrum of RCP (blue line) and LCP (red line) emissivity in the normal direction ($\theta=0,\phi=0$) as a function of wavelength. A large separation between these spectra leads to  $|S_T|\sim 1$ (black line) over a broad range of wavelengths. We note that $|S_T|$ is reasonably large beyond $\lambda \sim 16\mu $m as evident from the top figure. However, total emissivity given by $\eta_T = (\eta_{(\text{LCP})}+\eta_{(\text{RCP})})/2$ is extremely small at these higher wavelengths. Since the resulting thermal radiation power is extremely small, it will be challenging to detect CP light in potential experiments. Therefore, throughout this work, we only focus on the range of wavelengths where both $S_T$ and $\eta_T$ are reasonably large. The gyrotropy or the strength of nonreciprocity as quantified by the ratio  $|\varepsilon_{xy}|/|\varepsilon_{xx}| \gg 1$ over the same wavelength range is displayed by the middle figure. Both top and middle figures reveal the connection between strong gyrotropy and strong CP thermal radiation.
The bottom figure plots $S_T$ for all emission angles $\theta \in [0,\pi/2]$ for the same wavelength range. Because of the rotational symmetry due to perpendicular gyrotropy axis, these plots are the same for all in-plane angles $\phi \in [0,2\pi]$. The contour plot demonstrates that high purity ($|S_T|\sim 1$) CP light is emitted over not only a broad spectral bandwidth but also a broad range of emission angles ($\theta$). Based on the contour plot and the top figure, broadband high-purity CP thermal emission characterized by $|S_T| > 0.5$ and $\eta_T >0.5$, is realized over almost $75\%$ of the emission solid angle of $2\pi$ in the wavelength range of $6\mu$m to $14\mu$m for this representative MWS planar slab.

In Fig. 2b, we plot the same features for another well-studied nonreciprocal system of a doped InSb slab in magnetic field. For $B=1$T magnetic field applied perpendicular to InSb slab of large doping concentration $n=10^{17}\mathrm{cm}^{-3}$ (permittivity response taken from previous works
\cite{Chinmay2019}), we find that the emissivities of RCP and LCP light differ noticeably as shown in the top figure. Again, the connection with strong gyrotropy over the same wavelength range is evident from the middle figure. The broadband nature of $S_T$ (both emission directions and wavelengths) is depicted by the contour plot in the bottom figure. However, we note that this high purity thermal spin ($|S_T| \sim 1$) for InSb slab is obtained at wavelengths in the range of $\approx 40-100\mu$m. In typical experiments or practical applications, the operating temperatures can be considered to be in the range of $T=300$ to $1000$K. For such temperatures, the blackbody distribution  ($I_{\text{bb}}(\lambda,T)$) peaks at wavelengths in the range of $3-12\mu$m and decreases exponentially at larger wavelengths. The resulting low thermal emission power in addition to requirement of strong external magnetic field and limited availability of low-loss quarter wave plates over relevant wavelengths (needed for detection of nonzero $S_T$) are important requirements that make such doped semiconductors rather inconvenient for implementing the functionality of broadband CP thermal emission. These considerations are quite important from the perspective of high-precision experimental measurements of thermal radiation~\cite{xiao2020precision}. Other nonreciprocal media like ferro- and ferrimagnets exhibit large magnitude of $\varepsilon_{xy}$ in the wavelength range of $3-20\mu$m relevant for thermal radiation engineering. However, the gyrotropy remains small ($|\varepsilon_{xy}|/|\varepsilon_{xx}| \ll 1$) because of large negative permittivity ($\Re\{\varepsilon_{xx}\} \ll 0$). Furthermore, these materials also exhibit high reflectivity and low surface emissivity causing low thermal emission power. One can consider enhancing the gyrotropy in the relevant wavelength range based on patterning or resonant systems. However, such resonance-based approaches will necessarily lead to CP emission over narrow spectral and angular bandwidths. Optimization or inverse design of quarter wave metasurfaces (linear to CP conversion) or reciprocal chiral absorbers may be explored to obtain large bandwidths but no such work has been done so far. To the best of our knowledge, the bandwidths reported here using a representative MWS have not been realized before either theoretically or experimentally. Therefore, we believe that the planar slabs of MWS are uniquely capable of enabling the functionality of broadband CP thermal infrared radiation without patterning or application of external magnetic field.

\begin{figure*}[ht!]
\centering\includegraphics[width=0.88\linewidth]{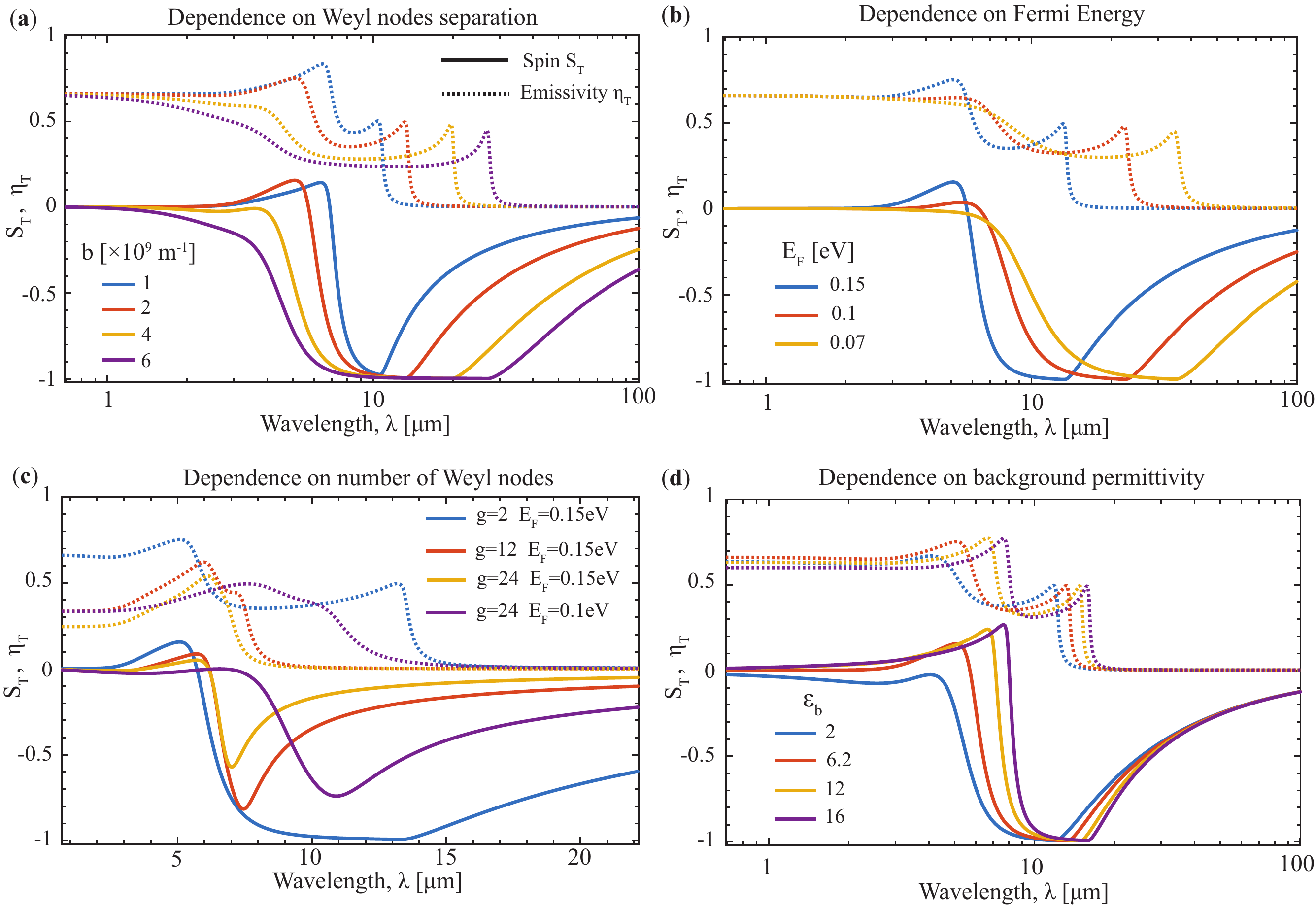}
\caption{The figure demonstrates the dependence of the broadband nature of CP thermal radiation in the normal emission direction ($\theta=0$) from the planar slab of MWS at $T=300$K on the underlying parameters of (a) Weyl nodes separation $b$, (b) Fermi energy $E_F$, (c) number of Weyl nodes $g$, and (d) background permittivity $\varepsilon_b$. The spectrum of thermal photon spin $S_T$ is shown by solid lines and that of the total emissivity $\eta_T$ is shown by dashed lines. (a) $b$ is varied at constant $E_F=0.15$eV, $g=2$, $\varepsilon_b=6.2$. (b) $E_F$ is varied at constant $b=2$nm$^{-1}$, $g=2$, $\varepsilon_b=6.2$. (c) $g$, $E_F$ are varied at constant $b=2$nm$^{-1}$, $\varepsilon_b=6.2$. (d) $\varepsilon_b$ is varied at constant $b=2$nm$^{-1}$, $E_F=0.15$eV, $g=2$.}
\label{fig3}
\end{figure*}

We now tune the underlying parameters in the model described in section~\ref{MWS} and analyze the behavior of thermal photon spin $S_T$ to point out favorable parameters for realizing broadband CP infrared radiation. We first vary the Weyl nodes separation $b$ in the above model along $\hat{\vk}_z$ from $1$ nm$^{-1}$ to 6 nm$^{-1}$.
Figure~\ref{fig3}(a) shows the spectrum of spin $S_T$ (solid lines) as well as of total emissivity $\eta_T$ (dashed lines) for increasing values of $b$. The same convention for $S_T$ and $\eta_T$ is then employed for all figures in Fig.~\ref{fig3}. Since the strength of nonreciprocity is directly proportional to Weyl nodes separation, the broadband nature of $S_T$ is further improved as the separation $b$ is increased. For instance, both $|S_T|\sim 0.5$ and reasonable emissivity $\eta_T\gtrsim 0.25$ are achieved over $4\mu$m to $28\mu$m wavelength range when $b=6$nm$^{-1}$ (violet lines).

Figure~\ref{fig3}(b) shows the dependence on Fermi energy by considering the values of $E_F=0.15,0.1,0.07$eV some of which are reported previously in \cite{Kotov2018,Co3Sn2S2,Co3Sn2S2_Co3Sn2Se2}. For these plots, the Weyl nodes separation is $b=2$nm$^{-1}$. As evident, with reduced Fermi energy, the bandwidth of CP thermal radiation increases and the spectrum shifts towards larger wavelengths because of smaller photon energies associated with the bands near low Fermi energy.

In the model of a representative MWS considered above, we used a simplified system containing only 2 Weyl nodes in the electronic band structure. This parameter, denoted as $g$, influences the diagonal permittivity of the Weyl semimetal through Eq. \ref{conductivity_eq}. Figure~\ref{fig3}(c) demonstrates the impact of $g$ on the thermal photon spin $S_T$. The blue, red, yellow plots corresponding to $g=2,12,24$ respectively are obtained when $b=2$nm$^{-1}$ and $E_F=0.15$eV. The resulting trend indicates that the bandwidth is reduced as the number of Weyl nodes is increased. The values of $g=12,24$ have been previously noted in Ref. \cite{Co3Sn2S2_Co3Sn2Se2,Sushkov2015}. In order to increase the bandwidth for a given number of Weyl nodes, we can either increase $b$ or decrease $E_F$ as illustrated in Figs.~\ref{fig3}(a,b). For example, comparison of violet ($g=24, E_F=0.1$eV) and yellow ($g=24, E_F=0.15$eV) lines in Fig.\ref{fig3}(c) indicates the feasibility of broadband CP light with larger number of Weyl nodes by reducing $E_F$.

Figure~\ref{fig3}(d) shows the dependence of the spectrum on the background permittivity denoted as $\varepsilon_b$. As evident, the spectrum slightly shifts towards larger wavelengths for increasing values of $\varepsilon_b$. Similar frequency shifts are observed for variations of reasonable magnitude in the other parameters in the model. For all results above, the temperature of the planar slab is assumed to be held at $T=300$K. When the temperature is increased to $T=400$K, small frequency shifts in the spectra are observed similar to those noted for total emissivity spectra in Ref.\cite{BoZhao2020}. Based on the numerical simulations, we find that the parameters $b$, $E_F$, $g$ are more impactful than other parameters ($\varepsilon_b, \nu_F,\tau, T$) considered in the permittivity model of MWS.

We also note that $S_T$ is quite small for the Voigt geometry in which the gyrotropy axis of MWS is parallel to the surface of the planar slab. This result was noted for InSb slab in Ref.~\cite{Chinmay2020} and was verified numerically for Weyl semimetal in this work. While Voigt geometry (parallel gyrotropy) is required for violation of directional Kirchhoff's law in other recent works \cite{BoZhao2020,Tsurimaki_Chen_Gang_2020}, Faraday geometry (perpendicular gyrotropy) is found to be more suitable for realizing strong CP thermal emission. Indeed, the Voigt geometry leads to very small thermal spin for the same magnitude of gyrotropy of the MWS in the Faraday geometry. Another important aspect concerns the partial linear polarization of thermal emission because of gyrotropy of the Weyl semimetal. Similar to the analysis of emissivities in the eigenbasis of RCP and LCP light, we can derive the emissivities in the eigenbasis of transverse electric ($s$) and transverse magnetic ($p$) polarized plane waves or in the eigenbasis of $s+p$ and $s-p$ polarized plane waves. We note that $\eta_{(s)} =\eta_{(p)}$ and $\eta_{(s+p)}=\eta_{(s-p)}$ for the normal emission direction ($\theta=0$) because of the rotational symmetry of the Faraday geometry considered above. At nonzero angles, we find that the differences in these emissivities ($|\eta_{(s)}-\eta_{(p)}|,|\eta_{(s+p)}-\eta_{(s-p)}|$) are very small compared to $|\eta_{\mathrm{(LCP)}}-\eta_{\mathrm{(RCP)}}|$. Thus, the gyrotropy or nonreciprocity of the Weyl semimetal in Faraday geometry is primarily connected with the spin angular momentum or circular polarization of emitted thermal photons. Because of these reasons, the analysis of nontrivial partial linear polarization for $\theta \neq 0$ emission directions in Voigt and Faraday geometries, which is rather straightforward, is not provided in this work.

\subsection{Finite subwavelength and wavelength-scale objects}
\label{finite}

In addition to extended planar slabs, we now show that the finite objects made of a magnetic Weyl semimetal can also emit broadband CP thermal radiation in specific emission directions. Again, there is no local formulation of Kirchhoff's law \cite{Greffet2018} which can simplify the calculation of emission by invoking electromagnetic reciprocity. Our system is nonreciprocal and the numerical calculation of circular polarization of thermal emission from these nonreciprocal bodies is more involved than that of reciprocal bodies. One numerical approach of analyzing thermal radiation from finite-size bodies is thermal discrete dipole approximation abbreviated as TDDA \cite{Edalatpour2014,TDDA_2017,Edalatpour2015}. Within TDDA, the finite body is subdivided into tiny subvolumes comprising of thermally fluctuating currents whose temperature-dependent strength is governed by the fluctuation-dissipation theorem (FDT). This approach was utilized to analyze radiative heat transfer in nonreciprocal systems \cite{TDDA_2017, Ott2018}, and was recently extended to analyze radiative angular momentum transfer \cite{Gao2020}. Using these recent developments, we calculate the thermal photon spin $S_T$ for thermal emission from a finite-size body as:
\begin{align}
S_T(\omega) &= \frac{\omega \inner{\vS(\omega)} }{ \inner{W(\omega)}} \cdot \hat{\vr} , \\
\inner{W(\omega)} &= \frac{1}{2}
(
\veps_0 \inner{\vE^*(\omega) \cdot \vE(\omega)} +
\mu_0   \inner{\vH^*(\omega) \cdot \vH(\omega)}
) ,
\\
\inner{\vS(\omega)} &= \frac{1}{2\omega}
\text{Im}(
\veps_0 \inner{\vE^*(\omega) \times \vE(\omega)} +
\mu_0   \inner{\vH^*(\omega) \times \vH(\omega)}
) ,
\end{align}
where $\inner{W(\omega)}$ denotes the spectral energy density and the vector $\inner{\vS(\omega)}$ denotes the spectral spin angular momentum density \cite{Bliokh2017,Barnett2016,yang2020quantum} at a given frequency $\omega$. $\hat{r}$ denotes the emission direction. This definition of spin in the context of thermal radiation has been previously explained in Ref.~\cite{chinmay_dipolar2019} and it quantifies the net spin angular momentum per emitted photon in the given emission direction ($\hat{r}$). In other words, the same quantity from the previous section is analyzed here. The quantities $\aver{\vE(\vr, \omega) \otimes \vE^*(\vr, \omega)}$ and $\aver{\vH(\vr, \omega) \otimes \vH^*(\vr, \omega)}$ are computed using the aformentioned TDDA approach applicable for finite-size nonreciprocal objects. From the perspective of potential experiments, it is also important to calculate the net emission power which has been described in several works before~\cite{TDDA_2017}. In the following, we plot the spectra of spin $S_T$ as well as the thermal emission power.

\begin{figure*}[ht!]
\centering\includegraphics[width=0.88\linewidth]{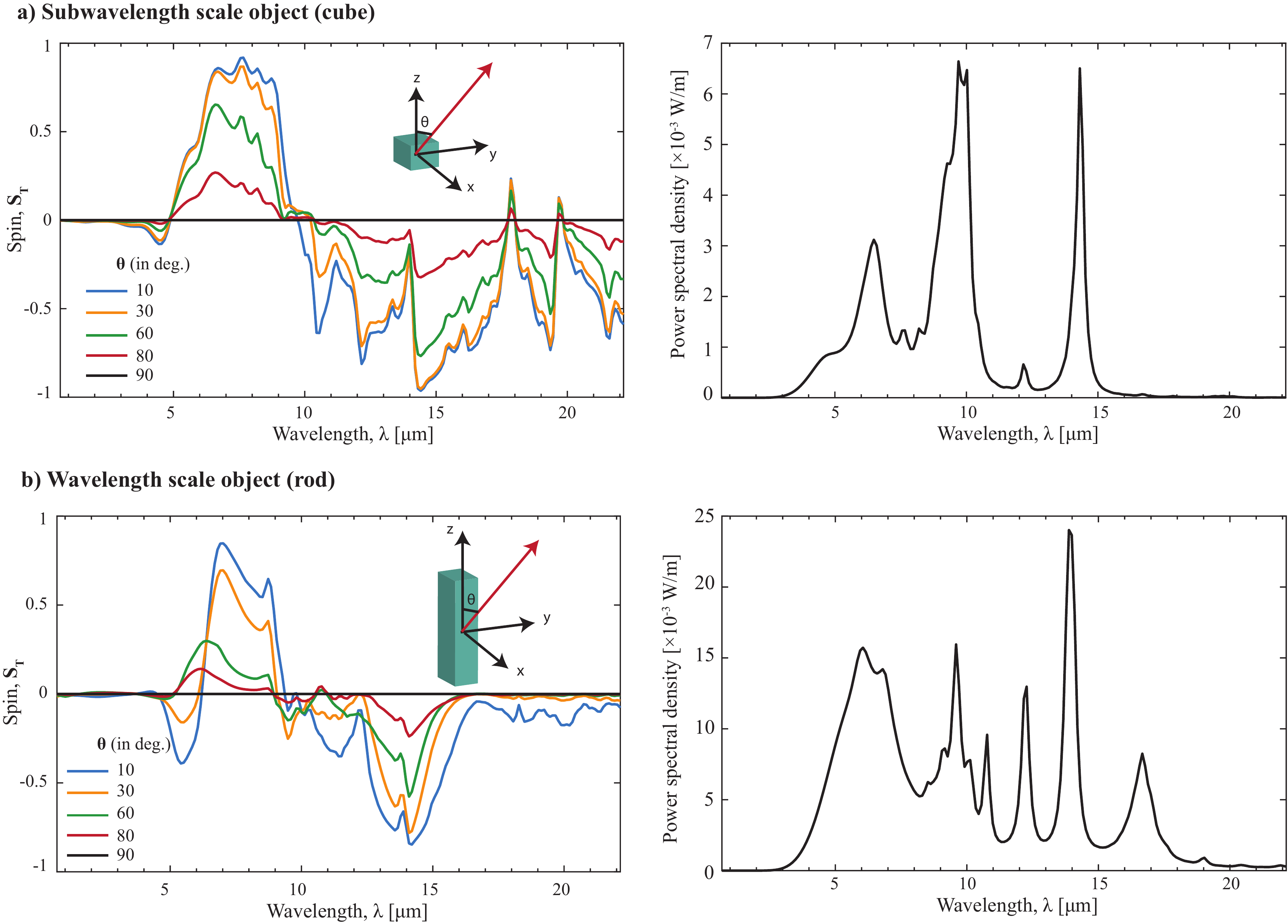}
\caption{Figure shows the circular polarization of thermal emission from (a) subwavelength-size cube and (b) wavelength-size rod of Weyl semimetal of parameters $b=2$nm$^{-1}$, $E_F=0.15$eV, $g=2$, $\varepsilon_b=6.2$ with gyrotopy axis along $\hat{z}$ direction. The dimensions of cube are $1\mu$m $\times 1\mu$m $\times 1\mu$m and the dimensions of the rod are $1\mu$m $\times 1\mu$m $\times 10\mu$m. The left figure plots the spectrum of thermal photon spin $S_T$ at various angles (emission directions) and the right figure plots the total angle-integrated power spectral density.
}
\label{fig4}
\end{figure*}

We first consider a cube made of MWS of length $1\mu$m along each side. The temperature of the cube is $T=300$K and that of its surrounding environment or vacuum is $T_0=0$K. We use the same representative MWS as used for the planar slab considered in Fig.~\ref{fig2} with gyrotropy axis along $\hat{z}$ direction. Using TDDA approach by discretizing the cube into $12\times 12\times 12$ subvolumes, we calculate $S_T$ in different emission directions in the far-field (over a sphere of radius $10^5\mu$m.). Numerical convergence in terms of less than $1\%$ relative change is observed upon changing the discretization scheme from $10\times 10\times 10$ subvolumes to $12\times 12\times 12$ subvolumes. As shown in the left figure of Fig.~\ref{fig4}(a), $S_T$ is plotted for emission directions characterized by $\theta \in [0,\pi]$. The dependence on the angle $\phi$ in the $xy$-plane is negligible because of the subwavelength size of the cube and the gyrotropy axis being perpendicular to the $xy$-plane. The spectrum of $S_T$ reveals the broadband partially CP thermal emission with $S_T$ being larger for smaller values of $\theta$. At $\theta=\pi/2$, $S_T=0$ at all wavelengths and for $\theta > \pi/2$, all the plots are flipped in sign ($S_T(\pi-\theta)=-S_T(\theta)$). The overall thermal emission power spectral density (integrated over all emission directions) is plotted in the right figure showing the wavelength range of interest from the perspective of potential experiments. In the wavelength range of interest between $5\mu$m to $15\mu$m where thermal emission power is dominant,  $|S_T|$ is close to one for a broad range of wavelengths in the emission direction characterized by $\theta$ close to either $0$ or $\pi$ values (parallel and antiparallel gyrotropy directions).
We also consider a rod of dimensions $1\mu$m $\times 1\mu$m $\times 10\mu$m with its longest side oriented along $\hat{z}$ direction. Numerical convergence of TDDA calculations is ensured in terms of less than $1\%$ relative change upon changing the discretization scheme from $8\times 8\times 80$ subvolumes to $10\times 10\times 100$ subvolumes. The resulting spectrum of spin $S_T$ calculated in the far field is plotted in the left figure of Fig.~\ref{fig4}(b) and the total thermal emission power spectral density is plotted in the right figure. The results are qualitatively similar to those displayed in Fig.~\ref{fig4}(a). Both figures prove the feasibility of broadband CP radiation from finite-size bodies of Weyl semimetals albeit with a smaller range of favorable emission directions compared to the extended planar slabs.

\section{Conclusion}
\label{conclusion}

We numerically demonstrated that planar slabs of magnetic Weyl semimetals (MWS) can emit high-purity CP thermal radiation ($S_T \sim 1$) over a broad wavelength range in mid-IR and long-wave-IR spectral region, and for a significant portion of the emission solid angle. For a representative MWS with parameters similar to Ref.~\cite{BoZhao2020}, strong circular polarization or thermal photon spin ($S_T > 0.5$) along with reasonably large emissivity ($\eta_T >0.3$) is realized over $6\mu$m to $14\mu$m wavelength range and $75\%$ portion of the emission solid angle of $2\pi$ for the planar slab. Based on theoretically well-established model of MWS, we analyzed the dependence of this result on underlying physical parameters and showed that even larger bandwidths can be achieved by tuning the favorable parameters. Specifically, larger separation between Weyl nodes (equivalent to stronger gyrotropy), smaller Fermi energy, and smaller number of Weyl nodes are favorable amongst other parameters for achieving larger bandwidths. We further demonstrated using TDDA computations that finite-size bodies made of MWS also lead to spectrally broadband CP thermal radiation albeit with geometry-dependent direction-specificity or small angular bandwidths compared to extended planar slabs.

We note that no such functionality of highly efficient broadband CP radiation at mid-IR and long-wave-IR wavelengths currently exists. It is difficult to realize simultaneously large spectral and angular bandwidths of high purity CP infrared emission reported in this work by employing other existing approaches. In other words, MWS appears to exhibit a unique capability to solve this technological design problem. There has been an extensive work in the past decade on Weyl semimetals with many candidates currently being investigated in experiments. In particular,  time-reversal-symmetry broken magnetic Weyl semimetals (focus of this work) include
pyrochlore iridates \ce{Y2IrO7}, \ce{Eu2IrO7}, \cite{Wan2011,Sushkov2015}
ferromagnetic spinels \ce{HgCr2Se4} \cite{Xu2011}, Heusler ferromagnets \ce{Co3Sn2S2},
\ce{Co3Sn2Se2} \cite{Co3Sn2S2_Co3Sn2Se2, Co3Sn2S2}, and Co-based Heusler alloy compounds \ce{Co2XZ} (X=V, Zr, Nb, Ti, Mn, Hf; Z=Si, Ge, Sn, Ga and Al)~\cite{kubler2012berry, chang2016room}
and ferrimagnetic \ce{Ti2MnAl}. Some of these candidate materials such as \ce{Co3Sn2S2}\cite{Liu_Co3Sn2S2_exp, Morali_Co3Sn2S2_exp}, \ce{Co2MnGa}\cite{Belopolski_Co2MnGa_exp}, \ce{Co2MnAl}~\cite{li2020giant} have been identified to be magnetic Weyl semimetals in recent experiments. Because of this great progress, we are confident that the central result of this work can be experimentally demonstrated soon.

\section*{Funding}
We acknowledge funding from DARPA Nascent Light-Matter Interactions program.

\section*{Disclosures}
The authors declare no conflicts of interest.


\section{Appendix}
\label{appendix}

We provide the summary of theoretical derivations and computational methods here.

\subsection{Spin resolved emissivity for extended planar slabs}

The polarization-resolved thermal radiation power from an extended planar slab is given by Eq.~\ref{Prad_planar} in the main text. This can be derived within the radiometry framework using detailed balance of energy and momenta or within the scattering formulation of fluctuational electrodynamics. The derivation for nonreciprocal media is discussed in detail in Ref.~\cite{Chinmay2020}. Here, we summarize the important details related to conventions of RCP and LCP light.

In the planar geometry, the emission direction is characterized by the angles $\theta,\phi$ made by the propagation wavevector $\hat{k}$. It is straightforward to obtain the associated transverse electric (s) and transverse magnetic (p) polarization unit vectors for the plane of incidence
spanned by $\hat{k}$ and $\hat{z}$ (normal to the slab surface). These eigenvectors are given below:
\begin{align}
    \hat{k} &= \bmat{\sin\theta \cos\phi \\ \sin\theta \sin\phi \\ \cos\theta},
    \ehat_s = \bmat{+\sin \phi \\ -\cos \phi \\ 0}, \nonumber \\
    \ehat_p &= \ehat_s \times \hat{k} = \bmat{-\cos \theta \cos \phi \\ -\cos \theta \sin \phi  \\ \sin \theta }.
\end{align}
The RCP eigenvector is $\ehat_s+i\ehat_p$ (also denoted as $+$) and the LCP eigenvector is $\ehat_s-i\ehat_p$ (also denoted as $-$). From the perspective of an observer/detector in the far-field, the incoming RCP photons have a clockwise rotation of $\mathbf{E}$-field and the incoming LCP photons have an anti-clockwise rotation of $\mathbf{E}$-field. The spin angular momentum per photon along the propagation direction is $\hbar$ for LCP photons and $-\hbar$ for RCP photons \cite{jackson1999}.

The emissivities in the eigenbasis of RCP and LCP photons are:
\begin{align}
\eta_{(\text{RCP})}&(\omega,\theta, \phi) = \eta_{(+)}(\omega,\theta, \phi) \nonumber\\
&=1-R_{(++)}(\omega,\theta, \phi+\pi)-R_{(+-)}(\omega, \theta, \phi+\pi), \\
\eta_{(\text{LCP})}&(\omega, \theta, \phi) = \eta_{(-)}(\omega, \theta, \phi) \nonumber\\
&=1-R_{(--)}(\omega, \theta, \phi+\pi)-R_{(-+)}(\omega, \theta, \phi+\pi).
\end{align}
where $R_{(ij)}(\omega, \theta, \phi)$ for $i,j \in \{+, -\}$ denotes the polarization interconversion reflectance for light of angular frequency $\omega$ incident in the direction characterized by the angles ($\theta,\phi$). These reflectances depend on the associated Fresnel reflection coefficients in $\ehat_s,\ehat_p$ basis and given by the following expressions:
\begin{align}
R_{(++/--)}&=\left|\left(r_{\mathrm{ss}}+r_{\mathrm{pp}}\right) \pm \mathrm{i}\left(r_{\mathrm{sp}}-r_{\mathrm{ps}}\right)\right|^{2} / 4, \\
R_{(-+/+-)}&=\left|\left(r_{\mathrm{ss}}-r_{\mathrm{pp}}\right) \pm \mathrm{i}\left(r_{\mathrm{sp}}+r_{\mathrm{ps}}\right)\right|^{2} / 4.
\end{align}
where we have omitted $(\omega, \theta, \phi)$-dependence for brevity since it is the same on both sides. $r_{jk}(\omega,\theta,\phi)$ denotes the amplitude of $j$-polarized reflected light due to unit-amplitude incident $k$-polarized light of frequency $\omega$. These reflection coefficients are computed by solving the boundary conditions. A computational tool to obtain these coefficients for generic bianisotropic media is available on GitHub \href{https://github.com/chinmayCK/Fresnel}{https://github.com/chinmayCK/Fresnel} under the MIT license.

\subsection{Thermal discrete dipole approximation approach}
We employ thermal discrete dipole approximation method recently used for analyzing thermal emission from nonreciprocal finite bodies \cite{TDDA_2017}. Here we clarify the assumptions and details in TDDA formalism specific to this work. Within TDDA formalism employed in this work, we discretize the finite-size objects into cubic sub-volumes containing point dipoles at their respective centers. Throughout this work, we assume the environment to be at $0$K. The electric field outside the object generated by discrete fluctuating point dipoles is expressed using dyadic Green's tensors as:
\begin{align}
    \vE(\vr) = \frac{k_0^2}{\veps_0} \sum_{i=1}^N \hat{G}_{EE} (\vr, \vr_i) \vp_i
    = \frac{k_0^2}{\veps_0} \GEE \vbP
\end{align}
where $\vbP$ is a column vector of length $3N$ consisting of dipoles $\vp_1, \dots, \vp_N$, and $\GEE = \bmat{\hat{G}_{EE}(\vr, \vr_1), \cdots, \hat{G}_{EE}(\vr, \vr_N)} $ is the row vector of
\begin{align}
    \hat{G}_{EE} (\vr, \vr') = \frac{e^{ik_0R}}{4\pi R}
    [&(1+\frac{i k_0 R-1}{k_0^2 R^2})\hat{1} + \nonumber\\
     &(
        \frac{3-3i k_0 R - k_0^2 R^2}{k_0^2 R^2})\frac{\vR \otimes \vR}{R^2}]
\end{align}
where $\vR = \vr-\vr', R = |\vr-\vr'|
$, and $\otimes$ denotes the outer product.
Here the bar notation of $\vbP$ and other following variables indicates that the quantities are located or evaluated inside the object.

The discrete dipoles inside the object denoted as $\vp_i$ consist of fluctuating part $\vp_i^{(\text{fl})}$ as well as the induced part $\vp_i^{(\text{ind})}$ which quantifies the dipole moments induced by external fields generated by other dipoles or the environment. This induced part can be expressed as a function of the exciting field as $\vp_i^{(\text{ind})}=\veps_0 \alpha_i \vE_i^{(\text{exc})}$ where $\alpha_i$ is the polarizability tensor of the discrete dipole.
\begin{align*}
    \vp_i &= \vp_i^{(\text{fl})} + \vp_i^{(\text{ind})} , \\
    \vp_i^{(\text{ind})} &= \veps_0 \alpha_i (
    \frac{k_0^2}{\veps_0} \sum_{j\neq i}^N \hat{G}_{EE} (\vr, \vr_j) \vp_j) .
\end{align*}
These equations in the matrix form are as the following:
\begin{align}
    \vbP &= \vbPfl + \vbPind , \nonumber \\
    \vbPind &= \veps_0 \bar{\alpha} (
    \frac{k_0^2}{\veps_0}\dGEE \vbP).
    \label{DDA}
\end{align}
This forms the discrete dipole approximation (DDA) equation which is to be solved to obtain $\vbP$ required to calculate thermal emission quantities.

The fluctuating point dipoles $\vp_i^{(\text{fl})}$ appearing in the above equations satisfy the fluctuation-dissipation theorem (FDT). For our calculations, the FDT leads to following expression for correlations in the matrix form:
\begin{equation}
    \aver{\vbPfl(\omega)\vbPfldagger(\omega')} = 2 \pi \hbar \veps_0 \delta(\omega-\omega')
    (1+2 n_{\mathrm{B}}(\omega, T)) \bar{\chi} 
\end{equation}
where $n_B(\omega, T) = 1/(e^{\hbar \omega / k_B T} - 1)$ is the Bose-Einstein distribution function, and $\bar{\chi} = \text{diag}(\hat{\chi}_1, \dots, \hat{\chi}_N)$, $\hat{\chi}_i = \hat{\chi}$ is a tensor combining the imaginary part of the polarizability tensor and the radiative correction,
\begin{align}
\hat{\chi} &= \frac{1}{2i}(\hat{\alpha} - \hat{\alpha}^{\dagger}) - \frac{k_0^3}{6\pi}, \hat{\alpha}^{\dagger}\hat{\alpha} .
\end{align}
And $\hat{\alpha} = (
\frac{1}{V_p}(\hat{L}_p + [\hat{\veps}_p - \mathbb{I}]^{-1})-i\frac{k_0^3}{6\pi}\mathbb{I}
)^{-1}$ is the polarizability tensor for cubic subvolume containng a material characterized by a generic permittivity tensor $\hat{\veps}_p$ which is not necessarily diagonal as in the case of isotropic materials. $V_p$ is the volume of the cubic element. $k_0=\omega/c$ where $c$ is speed of light and $\hat{L}_p = (1/3)\mathbb{I}$.

By solving this DDA equation given by Eq.\ref{DDA}, we get $\vbP = \mbT^{-1} \vbPfl$ where $\mbT = \mathbb{I} - k_0^2 \bar{\alpha} \dGEE$. Therefore,
\begin{align}
    \vE(\vr) = \frac{k_0^2}{\veps_0} \GEE (\mbT^{-1} \vbPfl) .
\end{align}
from which we compute the correlation $\aver{\vE(\vr, \omega) \otimes \vE^*(\vr, \omega)}$, which is
\begin{align}
\label{EE}
    \aver{\vE(\vr, \omega) \otimes \vE^*(\vr, \omega)} = \frac{k_0^4}{\veps_0^2}
    \GEE \mbT^{-1} \aver{\vbPfl\vbPfldagger} \mbT^{-1 \dagger} \GEE^\dagger .
\end{align}
Similarly, the magnetic field correlations $\aver{\vH(\vr, \omega) \otimes \vH^*(\vr, \omega)}$ are obtained by replacing the electric-electric dyadic Green's tensor $\hat{G}_{EE}$ by magnetic-electric tensor:
\begin{align}
    \hat{G}_{HE} (\vr, \vr') = \frac{e^{ik_0R}}{4\pi R}(1+\frac{i}{k_0 R}) \sqrt{\frac{\veps_0}{\mu_0}}
    \bmat{0 & -\vhr_z & \vhr_y \\
          \vhr_z & 0 & -\vhr_x \\
          -\vhr_y & \vhr_x & 0}
\end{align}
where $\vhr = \vR /R$. Thus we have
\begin{align}
\label{HH}
    \aver{\vH(\vr, \omega) \otimes \vH^*(\vr, \omega)} = \frac{k_0^4}{\veps_0^2}
    \GHE \mbT^{-1} \aver{\vbPfl\vbPfldagger} \mbT^{-1 \dagger} \GHE^\dagger .
\end{align}

Similarly, the correlations $\aver{\vE(\vr, \omega) \otimes \vH^*(\vr, \omega)}$ required for calculation of Poynting flux or radiative heat transfer can be  obtained within the TDDA formalism, which has been discussed in several works before~\cite{multiple_dipoles_TDDA_2013,Edalatpour2014,Edalatpour2015,TDDA_2017}. We employ the approach taken in Ref. \cite{TDDA_2017} in this work for calculating the angle-integrated thermal emission spectral density in the far field, plotted in Fig.\ref{fig4}.




\bibliography{sample}

\begin{thebibliography}{72}%
\makeatletter
\providecommand \@ifxundefined [1]{%
 \@ifx{#1\undefined}
}%
\providecommand \@ifnum [1]{%
 \ifnum #1\expandafter \@firstoftwo
 \else \expandafter \@secondoftwo
 \fi
}%
\providecommand \@ifx [1]{%
 \ifx #1\expandafter \@firstoftwo
 \else \expandafter \@secondoftwo
 \fi
}%
\providecommand \natexlab [1]{#1}%
\providecommand \enquote  [1]{``#1''}%
\providecommand \bibnamefont  [1]{#1}%
\providecommand \bibfnamefont [1]{#1}%
\providecommand \citenamefont [1]{#1}%
\providecommand \href@noop [0]{\@secondoftwo}%
\providecommand \href [0]{\begingroup \@sanitize@url \@href}%
\providecommand \@href[1]{\@@startlink{#1}\@@href}%
\providecommand \@@href[1]{\endgroup#1\@@endlink}%
\providecommand \@sanitize@url [0]{\catcode `\\12\catcode `\$12\catcode
  `\&12\catcode `\#12\catcode `\^12\catcode `\_12\catcode `\%12\relax}%
\providecommand \@@startlink[1]{}%
\providecommand \@@endlink[0]{}%
\providecommand \url  [0]{\begingroup\@sanitize@url \@url }%
\providecommand \@url [1]{\endgroup\@href {#1}{\urlprefix }}%
\providecommand \urlprefix  [0]{URL }%
\providecommand \Eprint [0]{\href }%
\providecommand \doibase [0]{https://doi.org/}%
\providecommand \selectlanguage [0]{\@gobble}%
\providecommand \bibinfo  [0]{\@secondoftwo}%
\providecommand \bibfield  [0]{\@secondoftwo}%
\providecommand \translation [1]{[#1]}%
\providecommand \BibitemOpen [0]{}%
\providecommand \bibitemStop [0]{}%
\providecommand \bibitemNoStop [0]{.\EOS\space}%
\providecommand \EOS [0]{\spacefactor3000\relax}%
\providecommand \BibitemShut  [1]{\csname bibitem#1\endcsname}%
\let\auto@bib@innerbib\@empty
\bibitem [{\citenamefont {Nishizawa}\ \emph {et~al.}(2017)\citenamefont
  {Nishizawa}, \citenamefont {Nishibayashi},\ and\ \citenamefont
  {Munekata}}]{Nishizawa2017}%
  \BibitemOpen
  \bibfield  {author} {\bibinfo {author} {\bibfnamefont {N.}~\bibnamefont
  {Nishizawa}}, \bibinfo {author} {\bibfnamefont {K.}~\bibnamefont
  {Nishibayashi}},\ and\ \bibinfo {author} {\bibfnamefont {H.}~\bibnamefont
  {Munekata}},\ }\bibfield  {title} {\bibinfo {title} {Pure circular
  polarization electroluminescence at room temperature with spin-polarized
  light-emitting diodes},\ }\href {https://doi.org/10.1073/pnas.1609839114}
  {\bibfield  {journal} {\bibinfo  {journal} {Proceedings of the National
  Academy of Sciences}\ }\textbf {\bibinfo {volume} {114}},\ \bibinfo {pages}
  {1783} (\bibinfo {year} {2017})}\BibitemShut {NoStop}%
\bibitem [{\citenamefont {Asshoff}\ \emph {et~al.}(2011)\citenamefont
  {Asshoff}, \citenamefont {Merz}, \citenamefont {Kalt},\ and\ \citenamefont
  {Hetterich}}]{Asshoff2011}%
  \BibitemOpen
  \bibfield  {author} {\bibinfo {author} {\bibfnamefont {P.}~\bibnamefont
  {Asshoff}}, \bibinfo {author} {\bibfnamefont {A.}~\bibnamefont {Merz}},
  \bibinfo {author} {\bibfnamefont {H.}~\bibnamefont {Kalt}},\ and\ \bibinfo
  {author} {\bibfnamefont {M.}~\bibnamefont {Hetterich}},\ }\bibfield  {title}
  {\bibinfo {title} {{A spintronic source of circularly polarized single
  photons}},\ }\href {https://doi.org/10.1063/1.3564893} {\bibfield  {journal}
  {\bibinfo  {journal} {Appl. Phys. Lett.}\ }\textbf {\bibinfo {volume} {98}},\
  \bibinfo {pages} {1} (\bibinfo {year} {2011})},\ \Eprint
  {https://arxiv.org/abs/1103.1117} {1103.1117} \BibitemShut {NoStop}%
\bibitem [{\citenamefont {{Di Nuzzo}}\ \emph {et~al.}(2017)\citenamefont {{Di
  Nuzzo}}, \citenamefont {Kulkarni}, \citenamefont {Zhao}, \citenamefont
  {Smolinsky}, \citenamefont {Tassinari}, \citenamefont {Meskers},
  \citenamefont {Naaman}, \citenamefont {Meijer},\ and\ \citenamefont
  {Friend}}]{DiNuzzo2017}%
  \BibitemOpen
  \bibfield  {author} {\bibinfo {author} {\bibfnamefont {D.}~\bibnamefont {{Di
  Nuzzo}}}, \bibinfo {author} {\bibfnamefont {C.}~\bibnamefont {Kulkarni}},
  \bibinfo {author} {\bibfnamefont {B.}~\bibnamefont {Zhao}}, \bibinfo {author}
  {\bibfnamefont {E.}~\bibnamefont {Smolinsky}}, \bibinfo {author}
  {\bibfnamefont {F.}~\bibnamefont {Tassinari}}, \bibinfo {author}
  {\bibfnamefont {S.~C.}\ \bibnamefont {Meskers}}, \bibinfo {author}
  {\bibfnamefont {R.}~\bibnamefont {Naaman}}, \bibinfo {author} {\bibfnamefont
  {E.~W.}\ \bibnamefont {Meijer}},\ and\ \bibinfo {author} {\bibfnamefont
  {R.~H.}\ \bibnamefont {Friend}},\ }\bibfield  {title} {\bibinfo {title}
  {{High Circular Polarization of Electroluminescence Achieved via
  Self-Assembly of a Light-Emitting Chiral Conjugated Polymer into Multidomain
  Cholesteric Films}},\ }\href {https://doi.org/10.1021/acsnano.7b07390}
  {\bibfield  {journal} {\bibinfo  {journal} {ACS Nano}\ }\textbf {\bibinfo
  {volume} {11}},\ \bibinfo {pages} {12713} (\bibinfo {year}
  {2017})}\BibitemShut {NoStop}%
\bibitem [{\citenamefont {Zhao}\ \emph {et~al.}(2016)\citenamefont {Zhao},
  \citenamefont {He}, \citenamefont {Gu}, \citenamefont {Guo}, \citenamefont
  {Wong}, \citenamefont {Lam},\ and\ \citenamefont {Tang}}]{Zhao2016}%
  \BibitemOpen
  \bibfield  {author} {\bibinfo {author} {\bibfnamefont {D.}~\bibnamefont
  {Zhao}}, \bibinfo {author} {\bibfnamefont {H.}~\bibnamefont {He}}, \bibinfo
  {author} {\bibfnamefont {X.}~\bibnamefont {Gu}}, \bibinfo {author}
  {\bibfnamefont {L.}~\bibnamefont {Guo}}, \bibinfo {author} {\bibfnamefont
  {K.~S.}\ \bibnamefont {Wong}}, \bibinfo {author} {\bibfnamefont {J.~W.~Y.}\
  \bibnamefont {Lam}},\ and\ \bibinfo {author} {\bibfnamefont {B.~Z.}\
  \bibnamefont {Tang}},\ }\bibfield  {title} {\bibinfo {title} {Circularly
  polarized luminescence and a reflective photoluminescent chiral nematic
  liquid crystal display based on an aggregation-induced emission luminogen},\
  }\href {https://doi.org/https://doi.org/10.1002/adom.201500646} {\bibfield
  {journal} {\bibinfo  {journal} {Advanced Optical Materials}\ }\textbf
  {\bibinfo {volume} {4}},\ \bibinfo {pages} {534} (\bibinfo {year}
  {2016})}\BibitemShut {NoStop}%
\bibitem [{\citenamefont {Zhang}\ \emph {et~al.}(2014)\citenamefont {Zhang},
  \citenamefont {Oka}, \citenamefont {Suzuki}, \citenamefont {Ye},\ and\
  \citenamefont {Iwasa}}]{Zhang2014}%
  \BibitemOpen
  \bibfield  {author} {\bibinfo {author} {\bibfnamefont {Y.~J.}\ \bibnamefont
  {Zhang}}, \bibinfo {author} {\bibfnamefont {T.}~\bibnamefont {Oka}}, \bibinfo
  {author} {\bibfnamefont {R.}~\bibnamefont {Suzuki}}, \bibinfo {author}
  {\bibfnamefont {J.~T.}\ \bibnamefont {Ye}},\ and\ \bibinfo {author}
  {\bibfnamefont {Y.}~\bibnamefont {Iwasa}},\ }\bibfield  {title} {\bibinfo
  {title} {{Electrically switchable chiral light-emitting transistor}},\ }\href
  {https://doi.org/10.1126/science.1251329} {\bibfield  {journal} {\bibinfo
  {journal} {Science (80-. ).}\ }\textbf {\bibinfo {volume} {344}},\ \bibinfo
  {pages} {725} (\bibinfo {year} {2014})}\BibitemShut {NoStop}%
\bibitem [{\citenamefont {Kumar}\ \emph {et~al.}(2015)\citenamefont {Kumar},
  \citenamefont {Nakashima},\ and\ \citenamefont {Kawai}}]{Kumar2015}%
  \BibitemOpen
  \bibfield  {author} {\bibinfo {author} {\bibfnamefont {J.}~\bibnamefont
  {Kumar}}, \bibinfo {author} {\bibfnamefont {T.}~\bibnamefont {Nakashima}},\
  and\ \bibinfo {author} {\bibfnamefont {T.}~\bibnamefont {Kawai}},\ }\bibfield
   {title} {\bibinfo {title} {{Circularly Polarized Luminescence in Chiral
  Molecules and Supramolecular Assemblies}},\ }\href
  {https://doi.org/10.1021/acs.jpclett.5b01452} {\bibfield  {journal} {\bibinfo
   {journal} {J. Phys. Chem. Lett.}\ }\textbf {\bibinfo {volume} {6}},\
  \bibinfo {pages} {3445} (\bibinfo {year} {2015})}\BibitemShut {NoStop}%
\bibitem [{\citenamefont {S{\'{a}}nchez-Carnerero}\ \emph
  {et~al.}(2015)\citenamefont {S{\'{a}}nchez-Carnerero}, \citenamefont
  {Agarrabeitia}, \citenamefont {Moreno}, \citenamefont {Maroto}, \citenamefont
  {Muller}, \citenamefont {Ortiz},\ and\ \citenamefont {{De La
  Moya}}}]{Sanchez-Carnerero2015}%
  \BibitemOpen
  \bibfield  {author} {\bibinfo {author} {\bibfnamefont {E.~M.}\ \bibnamefont
  {S{\'{a}}nchez-Carnerero}}, \bibinfo {author} {\bibfnamefont {A.~R.}\
  \bibnamefont {Agarrabeitia}}, \bibinfo {author} {\bibfnamefont
  {F.}~\bibnamefont {Moreno}}, \bibinfo {author} {\bibfnamefont {B.~L.}\
  \bibnamefont {Maroto}}, \bibinfo {author} {\bibfnamefont {G.}~\bibnamefont
  {Muller}}, \bibinfo {author} {\bibfnamefont {M.~J.}\ \bibnamefont {Ortiz}},\
  and\ \bibinfo {author} {\bibfnamefont {S.}~\bibnamefont {{De La Moya}}},\
  }\bibfield  {title} {\bibinfo {title} {{Circularly Polarized Luminescence
  from Simple Organic Molecules}},\ }\href
  {https://doi.org/10.1002/chem.201501178} {\bibfield  {journal} {\bibinfo
  {journal} {Chemistry - A European Journal}\ }\textbf {\bibinfo {volume}
  {21}},\ \bibinfo {pages} {13488} (\bibinfo {year} {2015})}\BibitemShut
  {NoStop}%
\bibitem [{\citenamefont {Konishi}\ \emph {et~al.}(2011)\citenamefont
  {Konishi}, \citenamefont {Nomura}, \citenamefont {Kumagai}, \citenamefont
  {Iwamoto}, \citenamefont {Arakawa},\ and\ \citenamefont
  {Kuwata-Gonokami}}]{Konishi2011}%
  \BibitemOpen
  \bibfield  {author} {\bibinfo {author} {\bibfnamefont {K.}~\bibnamefont
  {Konishi}}, \bibinfo {author} {\bibfnamefont {M.}~\bibnamefont {Nomura}},
  \bibinfo {author} {\bibfnamefont {N.}~\bibnamefont {Kumagai}}, \bibinfo
  {author} {\bibfnamefont {S.}~\bibnamefont {Iwamoto}}, \bibinfo {author}
  {\bibfnamefont {Y.}~\bibnamefont {Arakawa}},\ and\ \bibinfo {author}
  {\bibfnamefont {M.}~\bibnamefont {Kuwata-Gonokami}},\ }\bibfield  {title}
  {\bibinfo {title} {{Circularly polarized light emission from semiconductor
  planar chiral nanostructures}},\ }\href
  {https://doi.org/10.1103/PhysRevLett.106.057402} {\bibfield  {journal}
  {\bibinfo  {journal} {Phys. Rev. Lett.}\ }\textbf {\bibinfo {volume} {106}},\
  \bibinfo {pages} {1} (\bibinfo {year} {2011})}\BibitemShut {NoStop}%
\bibitem [{\citenamefont {Maksimov}\ \emph {et~al.}(2014)\citenamefont
  {Maksimov}, \citenamefont {Tartakovskii}, \citenamefont {Filatov},
  \citenamefont {Lobanov}, \citenamefont {Gippius}, \citenamefont {Tikhodeev},
  \citenamefont {Schneider}, \citenamefont {Kamp}, \citenamefont {Maier},
  \citenamefont {H\"ofling},\ and\ \citenamefont {Kulakovskii}}]{Maksimov2014}%
  \BibitemOpen
  \bibfield  {author} {\bibinfo {author} {\bibfnamefont {A.~A.}\ \bibnamefont
  {Maksimov}}, \bibinfo {author} {\bibfnamefont {I.~I.}\ \bibnamefont
  {Tartakovskii}}, \bibinfo {author} {\bibfnamefont {E.~V.}\ \bibnamefont
  {Filatov}}, \bibinfo {author} {\bibfnamefont {S.~V.}\ \bibnamefont
  {Lobanov}}, \bibinfo {author} {\bibfnamefont {N.~A.}\ \bibnamefont
  {Gippius}}, \bibinfo {author} {\bibfnamefont {S.~G.}\ \bibnamefont
  {Tikhodeev}}, \bibinfo {author} {\bibfnamefont {C.}~\bibnamefont
  {Schneider}}, \bibinfo {author} {\bibfnamefont {M.}~\bibnamefont {Kamp}},
  \bibinfo {author} {\bibfnamefont {S.}~\bibnamefont {Maier}}, \bibinfo
  {author} {\bibfnamefont {S.}~\bibnamefont {H\"ofling}},\ and\ \bibinfo
  {author} {\bibfnamefont {V.~D.}\ \bibnamefont {Kulakovskii}},\ }\bibfield
  {title} {\bibinfo {title} {Circularly polarized light emission from chiral
  spatially-structured planar semiconductor microcavities},\ }\href
  {https://doi.org/10.1103/PhysRevB.89.045316} {\bibfield  {journal} {\bibinfo
  {journal} {Phys. Rev. B}\ }\textbf {\bibinfo {volume} {89}},\ \bibinfo
  {pages} {045316} (\bibinfo {year} {2014})}\BibitemShut {NoStop}%
\bibitem [{\citenamefont {Wadsworth}\ \emph {et~al.}(2011)\citenamefont
  {Wadsworth}, \citenamefont {Clem}, \citenamefont {Branson},\ and\
  \citenamefont {Boreman}}]{Wadsworth2011}%
  \BibitemOpen
  \bibfield  {author} {\bibinfo {author} {\bibfnamefont {S.~L.}\ \bibnamefont
  {Wadsworth}}, \bibinfo {author} {\bibfnamefont {P.~G.}\ \bibnamefont {Clem}},
  \bibinfo {author} {\bibfnamefont {E.~D.}\ \bibnamefont {Branson}},\ and\
  \bibinfo {author} {\bibfnamefont {G.~D.}\ \bibnamefont {Boreman}},\
  }\bibfield  {title} {\bibinfo {title} {Broadband circularly-polarized
  infrared emission from multilayer metamaterials},\ }\href
  {https://doi.org/10.1364/OME.1.000466} {\bibfield  {journal} {\bibinfo
  {journal} {Opt. Mater. Express}\ }\textbf {\bibinfo {volume} {1}},\ \bibinfo
  {pages} {466} (\bibinfo {year} {2011})}\BibitemShut {NoStop}%
\bibitem [{\citenamefont {Shitrit}\ \emph {et~al.}(2013)\citenamefont
  {Shitrit}, \citenamefont {Yulevich}, \citenamefont {Maguid}, \citenamefont
  {Ozeri}, \citenamefont {Veksler}, \citenamefont {Kleiner},\ and\
  \citenamefont {Hasman}}]{Shitrit2013}%
  \BibitemOpen
  \bibfield  {author} {\bibinfo {author} {\bibfnamefont {N.}~\bibnamefont
  {Shitrit}}, \bibinfo {author} {\bibfnamefont {I.}~\bibnamefont {Yulevich}},
  \bibinfo {author} {\bibfnamefont {E.}~\bibnamefont {Maguid}}, \bibinfo
  {author} {\bibfnamefont {D.}~\bibnamefont {Ozeri}}, \bibinfo {author}
  {\bibfnamefont {D.}~\bibnamefont {Veksler}}, \bibinfo {author} {\bibfnamefont
  {V.}~\bibnamefont {Kleiner}},\ and\ \bibinfo {author} {\bibfnamefont
  {E.}~\bibnamefont {Hasman}},\ }\bibfield  {title} {\bibinfo {title}
  {Spin-optical metamaterial route to spin-controlled photonics},\ }\href
  {https://doi.org/10.1126/science.1234892} {\bibfield  {journal} {\bibinfo
  {journal} {Science}\ }\textbf {\bibinfo {volume} {340}},\ \bibinfo {pages}
  {724} (\bibinfo {year} {2013})}\BibitemShut {NoStop}%
\bibitem [{\citenamefont {Yang}\ and\ \citenamefont
  {Xu}(2010)}]{infrared_chiral_spectroscopy}%
  \BibitemOpen
  \bibfield  {author} {\bibinfo {author} {\bibfnamefont {G.}~\bibnamefont
  {Yang}}\ and\ \bibinfo {author} {\bibfnamefont {Y.}~\bibnamefont {Xu}},\
  }\bibfield  {title} {\bibinfo {title} {Vibrational circular dichroism
  spectroscopy of chiral molecules},\ }\href@noop {} {\bibfield  {journal}
  {\bibinfo  {journal} {Electronic and magnetic properties of chiral molecules
  and supramolecular architectures}\ ,\ \bibinfo {pages} {189}} (\bibinfo
  {year} {2010})}\BibitemShut {NoStop}%
\bibitem [{\citenamefont {Jasper}\ \emph {et~al.}(2020)\citenamefont {Jasper},
  \citenamefont {Mai}, \citenamefont {Warren}, \citenamefont {Smith},
  \citenamefont {Heligman}, \citenamefont {McCormick}, \citenamefont {Ou},
  \citenamefont {Sheffield},\ and\ \citenamefont
  {Vald\'es~Aguilar}}]{CP_spectroscopy}%
  \BibitemOpen
  \bibfield  {author} {\bibinfo {author} {\bibfnamefont {E.~V.}\ \bibnamefont
  {Jasper}}, \bibinfo {author} {\bibfnamefont {T.~T.}\ \bibnamefont {Mai}},
  \bibinfo {author} {\bibfnamefont {M.~T.}\ \bibnamefont {Warren}}, \bibinfo
  {author} {\bibfnamefont {R.~K.}\ \bibnamefont {Smith}}, \bibinfo {author}
  {\bibfnamefont {D.~M.}\ \bibnamefont {Heligman}}, \bibinfo {author}
  {\bibfnamefont {E.}~\bibnamefont {McCormick}}, \bibinfo {author}
  {\bibfnamefont {Y.~S.}\ \bibnamefont {Ou}}, \bibinfo {author} {\bibfnamefont
  {M.}~\bibnamefont {Sheffield}},\ and\ \bibinfo {author} {\bibfnamefont
  {R.}~\bibnamefont {Vald\'es~Aguilar}},\ }\bibfield  {title} {\bibinfo {title}
  {Broadband circular polarization time-domain terahertz spectroscopy},\ }\href
  {https://doi.org/10.1103/PhysRevMaterials.4.013803} {\bibfield  {journal}
  {\bibinfo  {journal} {Phys. Rev. Materials}\ }\textbf {\bibinfo {volume}
  {4}},\ \bibinfo {pages} {013803} (\bibinfo {year} {2020})}\BibitemShut
  {NoStop}%
\bibitem [{\citenamefont {Ranjbar}\ and\ \citenamefont
  {Gill}(2009{\natexlab{a}})}]{ranjbar2009circular}%
  \BibitemOpen
  \bibfield  {author} {\bibinfo {author} {\bibfnamefont {B.}~\bibnamefont
  {Ranjbar}}\ and\ \bibinfo {author} {\bibfnamefont {P.}~\bibnamefont {Gill}},\
  }\bibfield  {title} {\bibinfo {title} {Circular dichroism techniques:
  biomolecular and nanostructural analyses-a review},\ }\href@noop {}
  {\bibfield  {journal} {\bibinfo  {journal} {Chemical biology \& drug design}\
  }\textbf {\bibinfo {volume} {74}},\ \bibinfo {pages} {101} (\bibinfo {year}
  {2009}{\natexlab{a}})}\BibitemShut {NoStop}%
\bibitem [{\citenamefont {Basiri}\ \emph {et~al.}(2019)\citenamefont {Basiri},
  \citenamefont {Chen}, \citenamefont {Bai}, \citenamefont {Amrollahi},
  \citenamefont {Carpenter}, \citenamefont {Holman}, \citenamefont {Wang},\
  and\ \citenamefont {Yao}}]{basiri2019nature}%
  \BibitemOpen
  \bibfield  {author} {\bibinfo {author} {\bibfnamefont {A.}~\bibnamefont
  {Basiri}}, \bibinfo {author} {\bibfnamefont {X.}~\bibnamefont {Chen}},
  \bibinfo {author} {\bibfnamefont {J.}~\bibnamefont {Bai}}, \bibinfo {author}
  {\bibfnamefont {P.}~\bibnamefont {Amrollahi}}, \bibinfo {author}
  {\bibfnamefont {J.}~\bibnamefont {Carpenter}}, \bibinfo {author}
  {\bibfnamefont {Z.}~\bibnamefont {Holman}}, \bibinfo {author} {\bibfnamefont
  {C.}~\bibnamefont {Wang}},\ and\ \bibinfo {author} {\bibfnamefont
  {Y.}~\bibnamefont {Yao}},\ }\bibfield  {title} {\bibinfo {title}
  {{Nature-inspired chiral metasurfaces for circular polarization detection and
  full-Stokes polarimetric measurements}},\ }\href@noop {} {\bibfield
  {journal} {\bibinfo  {journal} {Light: Science \& Applications}\ }\textbf
  {\bibinfo {volume} {8}},\ \bibinfo {pages} {1} (\bibinfo {year}
  {2019})}\BibitemShut {NoStop}%
\bibitem [{\citenamefont {Hildebrand}\ \emph {et~al.}(2000)\citenamefont
  {Hildebrand}, \citenamefont {Davidson}, \citenamefont {Dotson}, \citenamefont
  {Dowell}, \citenamefont {Novak},\ and\ \citenamefont
  {Vaillancourt}}]{Hildebrand2000}%
  \BibitemOpen
  \bibfield  {author} {\bibinfo {author} {\bibfnamefont {R.}~\bibnamefont
  {Hildebrand}}, \bibinfo {author} {\bibfnamefont {J.}~\bibnamefont
  {Davidson}}, \bibinfo {author} {\bibfnamefont {J.}~\bibnamefont {Dotson}},
  \bibinfo {author} {\bibfnamefont {C.}~\bibnamefont {Dowell}}, \bibinfo
  {author} {\bibfnamefont {G.}~\bibnamefont {Novak}},\ and\ \bibinfo {author}
  {\bibfnamefont {J.}~\bibnamefont {Vaillancourt}},\ }\bibfield  {title}
  {\bibinfo {title} {{A Primer on Far‐Infrared Polarimetry}},\ }\href
  {https://doi.org/10.1086/316613} {\bibfield  {journal} {\bibinfo  {journal}
  {Publ. Astron. Soc. Pacific}\ }\textbf {\bibinfo {volume} {112}},\ \bibinfo
  {pages} {1215} (\bibinfo {year} {2000})}\BibitemShut {NoStop}%
\bibitem [{\citenamefont {Gade}\ and\ \citenamefont
  {Moeslund}(2014)}]{Gade2014}%
  \BibitemOpen
  \bibfield  {author} {\bibinfo {author} {\bibfnamefont {R.}~\bibnamefont
  {Gade}}\ and\ \bibinfo {author} {\bibfnamefont {T.~B.}\ \bibnamefont
  {Moeslund}},\ }\bibfield  {title} {\bibinfo {title} {{Thermal cameras and
  applications: A survey}},\ }\href {https://doi.org/10.1007/s00138-013-0570-5}
  {\bibfield  {journal} {\bibinfo  {journal} {Mach. Vis. Appl.}\ }\textbf
  {\bibinfo {volume} {25}},\ \bibinfo {pages} {245} (\bibinfo {year}
  {2014})}\BibitemShut {NoStop}%
\bibitem [{\citenamefont {Snik}\ \emph {et~al.}(2014)\citenamefont {Snik},
  \citenamefont {Craven-Jones}, \citenamefont {Escuti}, \citenamefont
  {Fineschi}, \citenamefont {Harrington}, \citenamefont {De~Martino},
  \citenamefont {Mawet}, \citenamefont {Riedi},\ and\ \citenamefont
  {Tyo}}]{snik2014overview}%
  \BibitemOpen
  \bibfield  {author} {\bibinfo {author} {\bibfnamefont {F.}~\bibnamefont
  {Snik}}, \bibinfo {author} {\bibfnamefont {J.}~\bibnamefont {Craven-Jones}},
  \bibinfo {author} {\bibfnamefont {M.}~\bibnamefont {Escuti}}, \bibinfo
  {author} {\bibfnamefont {S.}~\bibnamefont {Fineschi}}, \bibinfo {author}
  {\bibfnamefont {D.}~\bibnamefont {Harrington}}, \bibinfo {author}
  {\bibfnamefont {A.}~\bibnamefont {De~Martino}}, \bibinfo {author}
  {\bibfnamefont {D.}~\bibnamefont {Mawet}}, \bibinfo {author} {\bibfnamefont
  {J.}~\bibnamefont {Riedi}},\ and\ \bibinfo {author} {\bibfnamefont {J.~S.}\
  \bibnamefont {Tyo}},\ }\bibfield  {title} {\bibinfo {title} {An overview of
  polarimetric sensing techniques and technology with applications to different
  research fields},\ }in\ \href@noop {} {\emph {\bibinfo {booktitle}
  {Polarization: measurement, analysis, and remote sensing XI}}},\ Vol.\
  \bibinfo {volume} {9099}\ (\bibinfo {organization} {International Society for
  Optics and Photonics},\ \bibinfo {year} {2014})\ p.\ \bibinfo {pages}
  {90990B}\BibitemShut {NoStop}%
\bibitem [{\citenamefont {Bimonte}\ \emph {et~al.}(2009)\citenamefont
  {Bimonte}, \citenamefont {Cappellin}, \citenamefont {Carugno}, \citenamefont
  {Ruoso},\ and\ \citenamefont {Saadeh}}]{Bimonte2009}%
  \BibitemOpen
  \bibfield  {author} {\bibinfo {author} {\bibfnamefont {G.}~\bibnamefont
  {Bimonte}}, \bibinfo {author} {\bibfnamefont {L.}~\bibnamefont {Cappellin}},
  \bibinfo {author} {\bibfnamefont {G.}~\bibnamefont {Carugno}}, \bibinfo
  {author} {\bibfnamefont {G.}~\bibnamefont {Ruoso}},\ and\ \bibinfo {author}
  {\bibfnamefont {D.}~\bibnamefont {Saadeh}},\ }\bibfield  {title} {\bibinfo
  {title} {{Polarized thermal emission by thin metal wires}},\ }\bibfield
  {journal} {\bibinfo  {journal} {New J. Phys.}\ }\textbf {\bibinfo {volume}
  {11}},\ \href {https://doi.org/10.1088/1367-2630/11/3/033014}
  {10.1088/1367-2630/11/3/033014} (\bibinfo {year} {2009})\BibitemShut
  {NoStop}%
\bibitem [{\citenamefont {Marquier}\ \emph {et~al.}(2008)\citenamefont
  {Marquier}, \citenamefont {Arnold}, \citenamefont {Laroche}, \citenamefont
  {Greffet},\ and\ \citenamefont {Chen}}]{Marquier2008}%
  \BibitemOpen
  \bibfield  {author} {\bibinfo {author} {\bibfnamefont {F.}~\bibnamefont
  {Marquier}}, \bibinfo {author} {\bibfnamefont {C.}~\bibnamefont {Arnold}},
  \bibinfo {author} {\bibfnamefont {M.}~\bibnamefont {Laroche}}, \bibinfo
  {author} {\bibfnamefont {J.~J.}\ \bibnamefont {Greffet}},\ and\ \bibinfo
  {author} {\bibfnamefont {Y.}~\bibnamefont {Chen}},\ }\bibfield  {title}
  {\bibinfo {title} {{Degree of polarization of thermal light emitted by
  gratings supporting surface waves}},\ }\href
  {https://doi.org/10.1364/oe.16.005305} {\bibfield  {journal} {\bibinfo
  {journal} {Opt. Express}\ }\textbf {\bibinfo {volume} {16}},\ \bibinfo
  {pages} {5305} (\bibinfo {year} {2008})}\BibitemShut {NoStop}%
\bibitem [{\citenamefont {Dahan}\ \emph {et~al.}(2010)\citenamefont {Dahan},
  \citenamefont {Gorodetski}, \citenamefont {Frischwasser}, \citenamefont
  {Kleiner},\ and\ \citenamefont {Hasman}}]{Hasman_group_2010}%
  \BibitemOpen
  \bibfield  {author} {\bibinfo {author} {\bibfnamefont {N.}~\bibnamefont
  {Dahan}}, \bibinfo {author} {\bibfnamefont {Y.}~\bibnamefont {Gorodetski}},
  \bibinfo {author} {\bibfnamefont {K.}~\bibnamefont {Frischwasser}}, \bibinfo
  {author} {\bibfnamefont {V.}~\bibnamefont {Kleiner}},\ and\ \bibinfo {author}
  {\bibfnamefont {E.}~\bibnamefont {Hasman}},\ }\bibfield  {title} {\bibinfo
  {title} {Geometric doppler effect: Spin-split dispersion of thermal
  radiation},\ }\href {https://doi.org/10.1103/PhysRevLett.105.136402}
  {\bibfield  {journal} {\bibinfo  {journal} {Phys. Rev. Lett.}\ }\textbf
  {\bibinfo {volume} {105}},\ \bibinfo {pages} {136402} (\bibinfo {year}
  {2010})}\BibitemShut {NoStop}%
\bibitem [{\citenamefont {Yin}\ \emph {et~al.}(2013)\citenamefont {Yin},
  \citenamefont {Sch{\"{a}}ferling}, \citenamefont {Metzger},\ and\
  \citenamefont {Giessen}}]{Yin2013}%
  \BibitemOpen
  \bibfield  {author} {\bibinfo {author} {\bibfnamefont {X.}~\bibnamefont
  {Yin}}, \bibinfo {author} {\bibfnamefont {M.}~\bibnamefont
  {Sch{\"{a}}ferling}}, \bibinfo {author} {\bibfnamefont {B.}~\bibnamefont
  {Metzger}},\ and\ \bibinfo {author} {\bibfnamefont {H.}~\bibnamefont
  {Giessen}},\ }\bibfield  {title} {\bibinfo {title} {{Interpreting chiral
  nanophotonic spectra: The plasmonic Born-Kuhn model}},\ }\href
  {https://doi.org/10.1021/nl403705k} {\bibfield  {journal} {\bibinfo
  {journal} {Nano Lett.}\ }\textbf {\bibinfo {volume} {13}},\ \bibinfo {pages}
  {6238} (\bibinfo {year} {2013})}\BibitemShut {NoStop}%
\bibitem [{\citenamefont {Dyakov}\ \emph {et~al.}(2018)\citenamefont {Dyakov},
  \citenamefont {Semenenko}, \citenamefont {Gippius},\ and\ \citenamefont
  {Tikhodeev}}]{Dyakov2018}%
  \BibitemOpen
  \bibfield  {author} {\bibinfo {author} {\bibfnamefont {S.~A.}\ \bibnamefont
  {Dyakov}}, \bibinfo {author} {\bibfnamefont {V.~A.}\ \bibnamefont
  {Semenenko}}, \bibinfo {author} {\bibfnamefont {N.~A.}\ \bibnamefont
  {Gippius}},\ and\ \bibinfo {author} {\bibfnamefont {S.~G.}\ \bibnamefont
  {Tikhodeev}},\ }\bibfield  {title} {\bibinfo {title} {{Magnetic field free
  circularly polarized thermal emission from a chiral metasurface}},\ }\href
  {https://doi.org/10.1103/PhysRevB.98.235416} {\bibfield  {journal} {\bibinfo
  {journal} {Phys. Rev. B}\ }\textbf {\bibinfo {volume} {98}},\ \bibinfo
  {pages} {1} (\bibinfo {year} {2018})},\ \Eprint
  {https://arxiv.org/abs/1807.01752} {1807.01752} \BibitemShut {NoStop}%
\bibitem [{\citenamefont {Lee}\ and\ \citenamefont {Chan}(2007)}]{Lee2007}%
  \BibitemOpen
  \bibfield  {author} {\bibinfo {author} {\bibfnamefont {J.~C.~W.}\
  \bibnamefont {Lee}}\ and\ \bibinfo {author} {\bibfnamefont {C.~T.}\
  \bibnamefont {Chan}},\ }\bibfield  {title} {\bibinfo {title} {{Circularly
  polarized thermal radiation from layer-by-layer photonic crystal
  structures}},\ }\bibfield  {journal} {\bibinfo  {journal} {Appl. Phys.
  Lett.}\ }\textbf {\bibinfo {volume} {90}},\ \href
  {https://doi.org/10.1063/1.2435958} {10.1063/1.2435958} (\bibinfo {year}
  {2007})\BibitemShut {NoStop}%
\bibitem [{\citenamefont {Khandekar}\ and\ \citenamefont
  {Jacob}(2019{\natexlab{a}})}]{chinmay_dipolar2019}%
  \BibitemOpen
  \bibfield  {author} {\bibinfo {author} {\bibfnamefont {C.}~\bibnamefont
  {Khandekar}}\ and\ \bibinfo {author} {\bibfnamefont {Z.}~\bibnamefont
  {Jacob}},\ }\bibfield  {title} {\bibinfo {title} {{Circularly Polarized
  Thermal Radiation From Nonequilibrium Coupled Antennas}},\ }\href
  {https://doi.org/10.1103/PhysRevApplied.12.014053} {\bibfield  {journal}
  {\bibinfo  {journal} {Phys. Rev. Appl.}\ }\textbf {\bibinfo {volume} {12}},\
  \bibinfo {pages} {014053} (\bibinfo {year} {2019}{\natexlab{a}})}\BibitemShut
  {NoStop}%
\bibitem [{\citenamefont {Maghrebi}\ \emph {et~al.}(2019)\citenamefont
  {Maghrebi}, \citenamefont {Gorshkov},\ and\ \citenamefont
  {Sau}}]{Maghrebi_2019}%
  \BibitemOpen
  \bibfield  {author} {\bibinfo {author} {\bibfnamefont {M.~F.}\ \bibnamefont
  {Maghrebi}}, \bibinfo {author} {\bibfnamefont {A.~V.}\ \bibnamefont
  {Gorshkov}},\ and\ \bibinfo {author} {\bibfnamefont {J.~D.}\ \bibnamefont
  {Sau}},\ }\bibfield  {title} {\bibinfo {title} {Fluctuation-induced torque on
  a topological insulator out of thermal equilibrium},\ }\href
  {https://doi.org/10.1103/PhysRevLett.123.055901} {\bibfield  {journal}
  {\bibinfo  {journal} {Phys. Rev. Lett.}\ }\textbf {\bibinfo {volume} {123}},\
  \bibinfo {pages} {055901} (\bibinfo {year} {2019})}\BibitemShut {NoStop}%
\bibitem [{\citenamefont {Khan}\ and\ \citenamefont
  {Narimanov}(2019)}]{Narimanov2019}%
  \BibitemOpen
  \bibfield  {author} {\bibinfo {author} {\bibfnamefont {E.}~\bibnamefont
  {Khan}}\ and\ \bibinfo {author} {\bibfnamefont {E.~E.}\ \bibnamefont
  {Narimanov}},\ }\bibfield  {title} {\bibinfo {title} {Spinning radiation from
  a topological insulator},\ }\href
  {https://doi.org/10.1103/PhysRevB.100.081408} {\bibfield  {journal} {\bibinfo
   {journal} {Phys. Rev. B}\ }\textbf {\bibinfo {volume} {100}},\ \bibinfo
  {pages} {081408} (\bibinfo {year} {2019})}\BibitemShut {NoStop}%
\bibitem [{\citenamefont {Khandekar}\ \emph {et~al.}(2020)\citenamefont
  {Khandekar}, \citenamefont {Khosravi}, \citenamefont {Li},\ and\
  \citenamefont {Jacob}}]{Chinmay2020}%
  \BibitemOpen
  \bibfield  {author} {\bibinfo {author} {\bibfnamefont {C.}~\bibnamefont
  {Khandekar}}, \bibinfo {author} {\bibfnamefont {F.}~\bibnamefont {Khosravi}},
  \bibinfo {author} {\bibfnamefont {Z.}~\bibnamefont {Li}},\ and\ \bibinfo
  {author} {\bibfnamefont {Z.}~\bibnamefont {Jacob}},\ }\bibfield  {title}
  {\bibinfo {title} {{New spin-resolved thermal radiation laws for
  nonreciprocal bianisotropic media}},\ }\bibfield  {journal} {\bibinfo
  {journal} {New J. Phys.}\ }\textbf {\bibinfo {volume} {22}},\ \href
  {https://doi.org/10.1088/1367-2630/abc988} {10.1088/1367-2630/abc988}
  (\bibinfo {year} {2020})\BibitemShut {NoStop}%
\bibitem [{\citenamefont {Zhao}\ \emph {et~al.}(2020)\citenamefont {Zhao},
  \citenamefont {Guo}, \citenamefont {Garcia}, \citenamefont {Narang},\ and\
  \citenamefont {Fan}}]{BoZhao2020}%
  \BibitemOpen
  \bibfield  {author} {\bibinfo {author} {\bibfnamefont {B.}~\bibnamefont
  {Zhao}}, \bibinfo {author} {\bibfnamefont {C.}~\bibnamefont {Guo}}, \bibinfo
  {author} {\bibfnamefont {C.~A.~C.}\ \bibnamefont {Garcia}}, \bibinfo {author}
  {\bibfnamefont {P.}~\bibnamefont {Narang}},\ and\ \bibinfo {author}
  {\bibfnamefont {S.}~\bibnamefont {Fan}},\ }\bibfield  {title} {\bibinfo
  {title} {{Axion-Field-Enabled Nonreciprocal Thermal Radiation in Weyl
  Semimetals}},\ }\href {https://doi.org/10.1021/acs.nanolett.9b05179}
  {\bibfield  {journal} {\bibinfo  {journal} {Nano Lett.}\ }\textbf {\bibinfo
  {volume} {20}},\ \bibinfo {pages} {1923} (\bibinfo {year}
  {2020})}\BibitemShut {NoStop}%
\bibitem [{\citenamefont {Jia}\ \emph {et~al.}(2016)\citenamefont {Jia},
  \citenamefont {Xu},\ and\ \citenamefont {Hasan}}]{Jia2016}%
  \BibitemOpen
  \bibfield  {author} {\bibinfo {author} {\bibfnamefont {S.}~\bibnamefont
  {Jia}}, \bibinfo {author} {\bibfnamefont {S.~Y.}\ \bibnamefont {Xu}},\ and\
  \bibinfo {author} {\bibfnamefont {M.~Z.}\ \bibnamefont {Hasan}},\ }\bibfield
  {title} {\bibinfo {title} {{Weyl semimetals, Fermi arcs and chiral
  anomalies}},\ }\href {https://doi.org/10.1038/nmat4787} {\bibfield  {journal}
  {\bibinfo  {journal} {Nat. Mater.}\ }\textbf {\bibinfo {volume} {15}},\
  \bibinfo {pages} {1140} (\bibinfo {year} {2016})}\BibitemShut {NoStop}%
\bibitem [{\citenamefont {Huang}\ \emph {et~al.}(2015)\citenamefont {Huang},
  \citenamefont {Zhao}, \citenamefont {Long}, \citenamefont {Wang},
  \citenamefont {Chen}, \citenamefont {Yang}, \citenamefont {Liang},
  \citenamefont {Xue}, \citenamefont {Weng}, \citenamefont {Fang},
  \citenamefont {Dai},\ and\ \citenamefont {Chen}}]{Huang2015}%
  \BibitemOpen
  \bibfield  {author} {\bibinfo {author} {\bibfnamefont {X.}~\bibnamefont
  {Huang}}, \bibinfo {author} {\bibfnamefont {L.}~\bibnamefont {Zhao}},
  \bibinfo {author} {\bibfnamefont {Y.}~\bibnamefont {Long}}, \bibinfo {author}
  {\bibfnamefont {P.}~\bibnamefont {Wang}}, \bibinfo {author} {\bibfnamefont
  {D.}~\bibnamefont {Chen}}, \bibinfo {author} {\bibfnamefont {Z.}~\bibnamefont
  {Yang}}, \bibinfo {author} {\bibfnamefont {H.}~\bibnamefont {Liang}},
  \bibinfo {author} {\bibfnamefont {M.}~\bibnamefont {Xue}}, \bibinfo {author}
  {\bibfnamefont {H.}~\bibnamefont {Weng}}, \bibinfo {author} {\bibfnamefont
  {Z.}~\bibnamefont {Fang}}, \bibinfo {author} {\bibfnamefont {X.}~\bibnamefont
  {Dai}},\ and\ \bibinfo {author} {\bibfnamefont {G.}~\bibnamefont {Chen}},\
  }\bibfield  {title} {\bibinfo {title} {{Observation of the
  chiral-anomaly-induced negative magnetoresistance: In 3D Weyl semimetal
  TaAs}},\ }\href {https://doi.org/10.1103/PhysRevX.5.031023} {\bibfield
  {journal} {\bibinfo  {journal} {Phys. Rev. X}\ }\textbf {\bibinfo {volume}
  {5}},\ \bibinfo {pages} {1} (\bibinfo {year} {2015})},\ \Eprint
  {https://arxiv.org/abs/1503.01304} {1503.01304} \BibitemShut {NoStop}%
\bibitem [{\citenamefont {Wan}\ \emph {et~al.}(2011)\citenamefont {Wan},
  \citenamefont {Turner}, \citenamefont {Vishwanath},\ and\ \citenamefont
  {Savrasov}}]{Wan2011}%
  \BibitemOpen
  \bibfield  {author} {\bibinfo {author} {\bibfnamefont {X.}~\bibnamefont
  {Wan}}, \bibinfo {author} {\bibfnamefont {A.~M.}\ \bibnamefont {Turner}},
  \bibinfo {author} {\bibfnamefont {A.}~\bibnamefont {Vishwanath}},\ and\
  \bibinfo {author} {\bibfnamefont {S.~Y.}\ \bibnamefont {Savrasov}},\
  }\bibfield  {title} {\bibinfo {title} {Topological semimetal and fermi-arc
  surface states in the electronic structure of pyrochlore iridates},\ }\href
  {https://doi.org/10.1103/PhysRevB.83.205101} {\bibfield  {journal} {\bibinfo
  {journal} {Phys. Rev. B}\ }\textbf {\bibinfo {volume} {83}},\ \bibinfo
  {pages} {205101} (\bibinfo {year} {2011})}\BibitemShut {NoStop}%
\bibitem [{\citenamefont {Sushkov}\ \emph {et~al.}(2015)\citenamefont
  {Sushkov}, \citenamefont {Hofmann}, \citenamefont {Jenkins}, \citenamefont
  {Ishikawa}, \citenamefont {Nakatsuji}, \citenamefont {Das~Sarma},\ and\
  \citenamefont {Drew}}]{Sushkov2015}%
  \BibitemOpen
  \bibfield  {author} {\bibinfo {author} {\bibfnamefont {A.~B.}\ \bibnamefont
  {Sushkov}}, \bibinfo {author} {\bibfnamefont {J.~B.}\ \bibnamefont
  {Hofmann}}, \bibinfo {author} {\bibfnamefont {G.~S.}\ \bibnamefont
  {Jenkins}}, \bibinfo {author} {\bibfnamefont {J.}~\bibnamefont {Ishikawa}},
  \bibinfo {author} {\bibfnamefont {S.}~\bibnamefont {Nakatsuji}}, \bibinfo
  {author} {\bibfnamefont {S.}~\bibnamefont {Das~Sarma}},\ and\ \bibinfo
  {author} {\bibfnamefont {H.~D.}\ \bibnamefont {Drew}},\ }\bibfield  {title}
  {\bibinfo {title} {{Optical evidence for a Weyl semimetal state in pyrochlore
  $\mathrm{Eu_2Ir_2O_7}$ }},\ }\href
  {https://doi.org/10.1103/PhysRevB.92.241108} {\bibfield  {journal} {\bibinfo
  {journal} {Phys. Rev. B}\ }\textbf {\bibinfo {volume} {92}},\ \bibinfo
  {pages} {241108} (\bibinfo {year} {2015})}\BibitemShut {NoStop}%
\bibitem [{\citenamefont {Xu}\ \emph {et~al.}(2011)\citenamefont {Xu},
  \citenamefont {Weng}, \citenamefont {Wang}, \citenamefont {Dai},\ and\
  \citenamefont {Fang}}]{Xu2011}%
  \BibitemOpen
  \bibfield  {author} {\bibinfo {author} {\bibfnamefont {G.}~\bibnamefont
  {Xu}}, \bibinfo {author} {\bibfnamefont {H.}~\bibnamefont {Weng}}, \bibinfo
  {author} {\bibfnamefont {Z.}~\bibnamefont {Wang}}, \bibinfo {author}
  {\bibfnamefont {X.}~\bibnamefont {Dai}},\ and\ \bibinfo {author}
  {\bibfnamefont {Z.}~\bibnamefont {Fang}},\ }\bibfield  {title} {\bibinfo
  {title} {{Chern Semimetal and the Quantized Anomalous Hall Effect in
  {${\mathrm{HgCr}}_{2}{\mathrm{Se}}_{4}$}}},\ }\href
  {https://doi.org/10.1103/PhysRevLett.107.186806} {\bibfield  {journal}
  {\bibinfo  {journal} {Phys. Rev. Lett.}\ }\textbf {\bibinfo {volume} {107}},\
  \bibinfo {pages} {186806} (\bibinfo {year} {2011})}\BibitemShut {NoStop}%
\bibitem [{\citenamefont {K\"ubler}\ and\ \citenamefont
  {Felser}(2012)}]{kubler2012berry}%
  \BibitemOpen
  \bibfield  {author} {\bibinfo {author} {\bibfnamefont {J.}~\bibnamefont
  {K\"ubler}}\ and\ \bibinfo {author} {\bibfnamefont {C.}~\bibnamefont
  {Felser}},\ }\bibfield  {title} {\bibinfo {title} {Berry curvature and the
  anomalous hall effect in heusler compounds},\ }\href
  {https://doi.org/10.1103/PhysRevB.85.012405} {\bibfield  {journal} {\bibinfo
  {journal} {Phys. Rev. B}\ }\textbf {\bibinfo {volume} {85}},\ \bibinfo
  {pages} {012405} (\bibinfo {year} {2012})}\BibitemShut {NoStop}%
\bibitem [{\citenamefont {Chang}\ \emph {et~al.}(2016)\citenamefont {Chang},
  \citenamefont {Xu}, \citenamefont {Zheng}, \citenamefont {Singh},
  \citenamefont {Hsu}, \citenamefont {Bian}, \citenamefont {Alidoust},
  \citenamefont {Belopolski}, \citenamefont {Sanchez}, \citenamefont {Zhang}
  \emph {et~al.}}]{chang2016room}%
  \BibitemOpen
  \bibfield  {author} {\bibinfo {author} {\bibfnamefont {G.}~\bibnamefont
  {Chang}}, \bibinfo {author} {\bibfnamefont {S.-Y.}\ \bibnamefont {Xu}},
  \bibinfo {author} {\bibfnamefont {H.}~\bibnamefont {Zheng}}, \bibinfo
  {author} {\bibfnamefont {B.}~\bibnamefont {Singh}}, \bibinfo {author}
  {\bibfnamefont {C.-H.}\ \bibnamefont {Hsu}}, \bibinfo {author} {\bibfnamefont
  {G.}~\bibnamefont {Bian}}, \bibinfo {author} {\bibfnamefont {N.}~\bibnamefont
  {Alidoust}}, \bibinfo {author} {\bibfnamefont {I.}~\bibnamefont
  {Belopolski}}, \bibinfo {author} {\bibfnamefont {D.~S.}\ \bibnamefont
  {Sanchez}}, \bibinfo {author} {\bibfnamefont {S.}~\bibnamefont {Zhang}},
  \emph {et~al.},\ }\bibfield  {title} {\bibinfo {title} {{Room-temperature
  magnetic topological Weyl fermion and nodal line semimetal states in
  half-metallic Heusler $\mathrm{Co_2TiX}$ (X=Si, Ge, or Sn)}},\ }\href@noop {}
  {\bibfield  {journal} {\bibinfo  {journal} {Scientific reports}\ }\textbf
  {\bibinfo {volume} {6}},\ \bibinfo {pages} {1} (\bibinfo {year}
  {2016})}\BibitemShut {NoStop}%
\bibitem [{\citenamefont {Liu}\ \emph {et~al.}(2019)\citenamefont {Liu},
  \citenamefont {Liang}, \citenamefont {Liu}, \citenamefont {Xu}, \citenamefont
  {Li}, \citenamefont {Chen}, \citenamefont {Pei}, \citenamefont {Shi},
  \citenamefont {Mo}, \citenamefont {Dudin}, \citenamefont {Kim}, \citenamefont
  {Cacho}, \citenamefont {Li}, \citenamefont {Sun}, \citenamefont {Yang},
  \citenamefont {Liu}, \citenamefont {Parkin}, \citenamefont {Felser},\ and\
  \citenamefont {Chen}}]{Liu_Co3Sn2S2_exp}%
  \BibitemOpen
  \bibfield  {author} {\bibinfo {author} {\bibfnamefont {D.~F.}\ \bibnamefont
  {Liu}}, \bibinfo {author} {\bibfnamefont {A.~J.}\ \bibnamefont {Liang}},
  \bibinfo {author} {\bibfnamefont {E.~K.}\ \bibnamefont {Liu}}, \bibinfo
  {author} {\bibfnamefont {Q.~N.}\ \bibnamefont {Xu}}, \bibinfo {author}
  {\bibfnamefont {Y.~W.}\ \bibnamefont {Li}}, \bibinfo {author} {\bibfnamefont
  {C.}~\bibnamefont {Chen}}, \bibinfo {author} {\bibfnamefont {D.}~\bibnamefont
  {Pei}}, \bibinfo {author} {\bibfnamefont {W.~J.}\ \bibnamefont {Shi}},
  \bibinfo {author} {\bibfnamefont {S.~K.}\ \bibnamefont {Mo}}, \bibinfo
  {author} {\bibfnamefont {P.}~\bibnamefont {Dudin}}, \bibinfo {author}
  {\bibfnamefont {T.}~\bibnamefont {Kim}}, \bibinfo {author} {\bibfnamefont
  {C.}~\bibnamefont {Cacho}}, \bibinfo {author} {\bibfnamefont
  {G.}~\bibnamefont {Li}}, \bibinfo {author} {\bibfnamefont {Y.}~\bibnamefont
  {Sun}}, \bibinfo {author} {\bibfnamefont {L.~X.}\ \bibnamefont {Yang}},
  \bibinfo {author} {\bibfnamefont {Z.~K.}\ \bibnamefont {Liu}}, \bibinfo
  {author} {\bibfnamefont {S.~S.~P.}\ \bibnamefont {Parkin}}, \bibinfo {author}
  {\bibfnamefont {C.}~\bibnamefont {Felser}},\ and\ \bibinfo {author}
  {\bibfnamefont {Y.~L.}\ \bibnamefont {Chen}},\ }\bibfield  {title} {\bibinfo
  {title} {Magnetic weyl semimetal phase in a kagom{\'e} crystal},\ }\href
  {https://doi.org/10.1126/science.aav2873} {\bibfield  {journal} {\bibinfo
  {journal} {Science}\ }\textbf {\bibinfo {volume} {365}},\ \bibinfo {pages}
  {1282} (\bibinfo {year} {2019})}\BibitemShut {NoStop}%
\bibitem [{\citenamefont {Morali}\ \emph {et~al.}(2019)\citenamefont {Morali},
  \citenamefont {Batabyal}, \citenamefont {Nag}, \citenamefont {Liu},
  \citenamefont {Xu}, \citenamefont {Sun}, \citenamefont {Yan}, \citenamefont
  {Felser}, \citenamefont {Avraham},\ and\ \citenamefont
  {Beidenkopf}}]{Morali_Co3Sn2S2_exp}%
  \BibitemOpen
  \bibfield  {author} {\bibinfo {author} {\bibfnamefont {N.}~\bibnamefont
  {Morali}}, \bibinfo {author} {\bibfnamefont {R.}~\bibnamefont {Batabyal}},
  \bibinfo {author} {\bibfnamefont {P.~K.}\ \bibnamefont {Nag}}, \bibinfo
  {author} {\bibfnamefont {E.}~\bibnamefont {Liu}}, \bibinfo {author}
  {\bibfnamefont {Q.}~\bibnamefont {Xu}}, \bibinfo {author} {\bibfnamefont
  {Y.}~\bibnamefont {Sun}}, \bibinfo {author} {\bibfnamefont {B.}~\bibnamefont
  {Yan}}, \bibinfo {author} {\bibfnamefont {C.}~\bibnamefont {Felser}},
  \bibinfo {author} {\bibfnamefont {N.}~\bibnamefont {Avraham}},\ and\ \bibinfo
  {author} {\bibfnamefont {H.}~\bibnamefont {Beidenkopf}},\ }\bibfield  {title}
  {\bibinfo {title} {{Fermi-arc diversity on surface terminations of the
  magnetic Weyl semimetal Co3Sn2S2}},\ }\href
  {https://doi.org/10.1126/science.aav2334} {\bibfield  {journal} {\bibinfo
  {journal} {Science}\ }\textbf {\bibinfo {volume} {365}},\ \bibinfo {pages}
  {1286} (\bibinfo {year} {2019})}\BibitemShut {NoStop}%
\bibitem [{\citenamefont {Belopolski}\ \emph {et~al.}(2019)\citenamefont
  {Belopolski}, \citenamefont {Manna}, \citenamefont {Sanchez}, \citenamefont
  {Chang}, \citenamefont {Ernst}, \citenamefont {Yin}, \citenamefont {Zhang},
  \citenamefont {Cochran}, \citenamefont {Shumiya}, \citenamefont {Zheng},
  \citenamefont {Singh}, \citenamefont {Bian}, \citenamefont {Multer},
  \citenamefont {Litskevich}, \citenamefont {Zhou}, \citenamefont {Huang},
  \citenamefont {Wang}, \citenamefont {Chang}, \citenamefont {Xu},
  \citenamefont {Bansil}, \citenamefont {Felser}, \citenamefont {Lin},\ and\
  \citenamefont {Hasan}}]{Belopolski_Co2MnGa_exp}%
  \BibitemOpen
  \bibfield  {author} {\bibinfo {author} {\bibfnamefont {I.}~\bibnamefont
  {Belopolski}}, \bibinfo {author} {\bibfnamefont {K.}~\bibnamefont {Manna}},
  \bibinfo {author} {\bibfnamefont {D.~S.}\ \bibnamefont {Sanchez}}, \bibinfo
  {author} {\bibfnamefont {G.}~\bibnamefont {Chang}}, \bibinfo {author}
  {\bibfnamefont {B.}~\bibnamefont {Ernst}}, \bibinfo {author} {\bibfnamefont
  {J.}~\bibnamefont {Yin}}, \bibinfo {author} {\bibfnamefont {S.~S.}\
  \bibnamefont {Zhang}}, \bibinfo {author} {\bibfnamefont {T.}~\bibnamefont
  {Cochran}}, \bibinfo {author} {\bibfnamefont {N.}~\bibnamefont {Shumiya}},
  \bibinfo {author} {\bibfnamefont {H.}~\bibnamefont {Zheng}}, \bibinfo
  {author} {\bibfnamefont {B.}~\bibnamefont {Singh}}, \bibinfo {author}
  {\bibfnamefont {G.}~\bibnamefont {Bian}}, \bibinfo {author} {\bibfnamefont
  {D.}~\bibnamefont {Multer}}, \bibinfo {author} {\bibfnamefont
  {M.}~\bibnamefont {Litskevich}}, \bibinfo {author} {\bibfnamefont
  {X.}~\bibnamefont {Zhou}}, \bibinfo {author} {\bibfnamefont {S.-M.}\
  \bibnamefont {Huang}}, \bibinfo {author} {\bibfnamefont {B.}~\bibnamefont
  {Wang}}, \bibinfo {author} {\bibfnamefont {T.-R.}\ \bibnamefont {Chang}},
  \bibinfo {author} {\bibfnamefont {S.-Y.}\ \bibnamefont {Xu}}, \bibinfo
  {author} {\bibfnamefont {A.}~\bibnamefont {Bansil}}, \bibinfo {author}
  {\bibfnamefont {C.}~\bibnamefont {Felser}}, \bibinfo {author} {\bibfnamefont
  {H.}~\bibnamefont {Lin}},\ and\ \bibinfo {author} {\bibfnamefont {M.~Z.}\
  \bibnamefont {Hasan}},\ }\bibfield  {title} {\bibinfo {title} {Discovery of
  topological weyl fermion lines and drumhead surface states in a room
  temperature magnet},\ }\href {https://doi.org/10.1126/science.aav2327}
  {\bibfield  {journal} {\bibinfo  {journal} {Science}\ }\textbf {\bibinfo
  {volume} {365}},\ \bibinfo {pages} {1278} (\bibinfo {year}
  {2019})}\BibitemShut {NoStop}%
\bibitem [{\citenamefont {Li}\ \emph {et~al.}(2020)\citenamefont {Li},
  \citenamefont {Koo}, \citenamefont {Ning}, \citenamefont {Li}, \citenamefont
  {Miao}, \citenamefont {Min}, \citenamefont {Zhu}, \citenamefont {Wang},
  \citenamefont {Alem}, \citenamefont {Liu} \emph {et~al.}}]{li2020giant}%
  \BibitemOpen
  \bibfield  {author} {\bibinfo {author} {\bibfnamefont {P.}~\bibnamefont
  {Li}}, \bibinfo {author} {\bibfnamefont {J.}~\bibnamefont {Koo}}, \bibinfo
  {author} {\bibfnamefont {W.}~\bibnamefont {Ning}}, \bibinfo {author}
  {\bibfnamefont {J.}~\bibnamefont {Li}}, \bibinfo {author} {\bibfnamefont
  {L.}~\bibnamefont {Miao}}, \bibinfo {author} {\bibfnamefont {L.}~\bibnamefont
  {Min}}, \bibinfo {author} {\bibfnamefont {Y.}~\bibnamefont {Zhu}}, \bibinfo
  {author} {\bibfnamefont {Y.}~\bibnamefont {Wang}}, \bibinfo {author}
  {\bibfnamefont {N.}~\bibnamefont {Alem}}, \bibinfo {author} {\bibfnamefont
  {C.-X.}\ \bibnamefont {Liu}}, \emph {et~al.},\ }\bibfield  {title} {\bibinfo
  {title} {{Giant room temperature anomalous Hall effect and tunable topology
  in a ferromagnetic topological semimetal $\mathrm{Co_2MnAl}$}},\ }\href@noop
  {} {\bibfield  {journal} {\bibinfo  {journal} {Nature communications}\
  }\textbf {\bibinfo {volume} {11}},\ \bibinfo {pages} {1} (\bibinfo {year}
  {2020})}\BibitemShut {NoStop}%
\bibitem [{\citenamefont {Morimoto}\ \emph {et~al.}(2016)\citenamefont
  {Morimoto}, \citenamefont {Zhong}, \citenamefont {Orenstein},\ and\
  \citenamefont {Moore}}]{Morimoto2016}%
  \BibitemOpen
  \bibfield  {author} {\bibinfo {author} {\bibfnamefont {T.}~\bibnamefont
  {Morimoto}}, \bibinfo {author} {\bibfnamefont {S.}~\bibnamefont {Zhong}},
  \bibinfo {author} {\bibfnamefont {J.}~\bibnamefont {Orenstein}},\ and\
  \bibinfo {author} {\bibfnamefont {J.~E.}\ \bibnamefont {Moore}},\ }\bibfield
  {title} {\bibinfo {title} {Semiclassical theory of nonlinear magneto-optical
  responses with applications to topological dirac/weyl semimetals},\ }\href
  {https://doi.org/10.1103/PhysRevB.94.245121} {\bibfield  {journal} {\bibinfo
  {journal} {Phys. Rev. B}\ }\textbf {\bibinfo {volume} {94}},\ \bibinfo
  {pages} {245121} (\bibinfo {year} {2016})}\BibitemShut {NoStop}%
\bibitem [{\citenamefont {Osterhoudt}\ \emph {et~al.}(2019)\citenamefont
  {Osterhoudt}, \citenamefont {Diebel}, \citenamefont {Gray}, \citenamefont
  {Yang}, \citenamefont {Stanco}, \citenamefont {Huang}, \citenamefont {Shen},
  \citenamefont {Ni}, \citenamefont {Moll}, \citenamefont {Ran} \emph
  {et~al.}}]{WSM_photovoltaic}%
  \BibitemOpen
  \bibfield  {author} {\bibinfo {author} {\bibfnamefont {G.~B.}\ \bibnamefont
  {Osterhoudt}}, \bibinfo {author} {\bibfnamefont {L.~K.}\ \bibnamefont
  {Diebel}}, \bibinfo {author} {\bibfnamefont {M.~J.}\ \bibnamefont {Gray}},
  \bibinfo {author} {\bibfnamefont {X.}~\bibnamefont {Yang}}, \bibinfo {author}
  {\bibfnamefont {J.}~\bibnamefont {Stanco}}, \bibinfo {author} {\bibfnamefont
  {X.}~\bibnamefont {Huang}}, \bibinfo {author} {\bibfnamefont
  {B.}~\bibnamefont {Shen}}, \bibinfo {author} {\bibfnamefont {N.}~\bibnamefont
  {Ni}}, \bibinfo {author} {\bibfnamefont {P.~J.}\ \bibnamefont {Moll}},
  \bibinfo {author} {\bibfnamefont {Y.}~\bibnamefont {Ran}}, \emph {et~al.},\
  }\bibfield  {title} {\bibinfo {title} {Colossal mid-infrared bulk
  photovoltaic effect in a type-i weyl semimetal},\ }\href@noop {} {\bibfield
  {journal} {\bibinfo  {journal} {Nature materials}\ }\textbf {\bibinfo
  {volume} {18}},\ \bibinfo {pages} {471} (\bibinfo {year} {2019})}\BibitemShut
  {NoStop}%
\bibitem [{\citenamefont {de~Juan}\ \emph {et~al.}(2017)\citenamefont
  {de~Juan}, \citenamefont {Grushin}, \citenamefont {Morimoto},\ and\
  \citenamefont {Moore}}]{WSM_photogalvanic}%
  \BibitemOpen
  \bibfield  {author} {\bibinfo {author} {\bibfnamefont {F.}~\bibnamefont
  {de~Juan}}, \bibinfo {author} {\bibfnamefont {A.~G.}\ \bibnamefont
  {Grushin}}, \bibinfo {author} {\bibfnamefont {T.}~\bibnamefont {Morimoto}},\
  and\ \bibinfo {author} {\bibfnamefont {J.~E.}\ \bibnamefont {Moore}},\
  }\bibfield  {title} {\bibinfo {title} {Quantized circular photogalvanic
  effect in weyl semimetals},\ }\href@noop {} {\bibfield  {journal} {\bibinfo
  {journal} {Nature communications}\ }\textbf {\bibinfo {volume} {8}},\
  \bibinfo {pages} {1} (\bibinfo {year} {2017})}\BibitemShut {NoStop}%
\bibitem [{\citenamefont {Kotov}\ and\ \citenamefont
  {Lozovik}(2016)}]{Kotov2016}%
  \BibitemOpen
  \bibfield  {author} {\bibinfo {author} {\bibfnamefont {O.~V.}\ \bibnamefont
  {Kotov}}\ and\ \bibinfo {author} {\bibfnamefont {Y.~E.}\ \bibnamefont
  {Lozovik}},\ }\bibfield  {title} {\bibinfo {title} {Dielectric response and
  novel electromagnetic modes in three-dimensional dirac semimetal films},\
  }\href {https://doi.org/10.1103/PhysRevB.93.235417} {\bibfield  {journal}
  {\bibinfo  {journal} {Phys. Rev. B}\ }\textbf {\bibinfo {volume} {93}},\
  \bibinfo {pages} {235417} (\bibinfo {year} {2016})}\BibitemShut {NoStop}%
\bibitem [{\citenamefont {Kotov}\ and\ \citenamefont
  {Lozovik}(2018)}]{Kotov2018}%
  \BibitemOpen
  \bibfield  {author} {\bibinfo {author} {\bibfnamefont {O.~V.}\ \bibnamefont
  {Kotov}}\ and\ \bibinfo {author} {\bibfnamefont {Y.~E.}\ \bibnamefont
  {Lozovik}},\ }\bibfield  {title} {\bibinfo {title} {Giant tunable
  nonreciprocity of light in weyl semimetals},\ }\href
  {https://doi.org/10.1103/PhysRevB.98.195446} {\bibfield  {journal} {\bibinfo
  {journal} {Phys. Rev. B}\ }\textbf {\bibinfo {volume} {98}},\ \bibinfo
  {pages} {195446} (\bibinfo {year} {2018})}\BibitemShut {NoStop}%
\bibitem [{\citenamefont {Asadchy}\ \emph {et~al.}(2020)\citenamefont
  {Asadchy}, \citenamefont {Guo}, \citenamefont {Zhao},\ and\ \citenamefont
  {Fan}}]{Optical_Isolators}%
  \BibitemOpen
  \bibfield  {author} {\bibinfo {author} {\bibfnamefont {V.~S.}\ \bibnamefont
  {Asadchy}}, \bibinfo {author} {\bibfnamefont {C.}~\bibnamefont {Guo}},
  \bibinfo {author} {\bibfnamefont {B.}~\bibnamefont {Zhao}},\ and\ \bibinfo
  {author} {\bibfnamefont {S.}~\bibnamefont {Fan}},\ }\bibfield  {title}
  {\bibinfo {title} {Sub-wavelength passive optical isolators using photonic
  structures based on weyl semimetals},\ }\href
  {https://doi.org/https://doi.org/10.1002/adom.202000100} {\bibfield
  {journal} {\bibinfo  {journal} {Advanced Optical Materials}\ }\textbf
  {\bibinfo {volume} {8}},\ \bibinfo {pages} {2000100} (\bibinfo {year}
  {2020})}\BibitemShut {NoStop}%
\bibitem [{\citenamefont {Green}(2012)}]{Green_Martin_PV}%
  \BibitemOpen
  \bibfield  {author} {\bibinfo {author} {\bibfnamefont {M.~A.}\ \bibnamefont
  {Green}},\ }\bibfield  {title} {\bibinfo {title} {Time-asymmetric
  photovoltaics},\ }\href {https://doi.org/10.1021/nl3034784} {\bibfield
  {journal} {\bibinfo  {journal} {Nano Letters}\ }\textbf {\bibinfo {volume}
  {12}},\ \bibinfo {pages} {5985} (\bibinfo {year} {2012})}\BibitemShut
  {NoStop}%
\bibitem [{\citenamefont {Tsurimaki}\ \emph {et~al.}(2020)\citenamefont
  {Tsurimaki}, \citenamefont {Qian}, \citenamefont {Pajovic}, \citenamefont
  {Han}, \citenamefont {Li},\ and\ \citenamefont
  {Chen}}]{Tsurimaki_Chen_Gang_2020}%
  \BibitemOpen
  \bibfield  {author} {\bibinfo {author} {\bibfnamefont {Y.}~\bibnamefont
  {Tsurimaki}}, \bibinfo {author} {\bibfnamefont {X.}~\bibnamefont {Qian}},
  \bibinfo {author} {\bibfnamefont {S.}~\bibnamefont {Pajovic}}, \bibinfo
  {author} {\bibfnamefont {F.}~\bibnamefont {Han}}, \bibinfo {author}
  {\bibfnamefont {M.}~\bibnamefont {Li}},\ and\ \bibinfo {author}
  {\bibfnamefont {G.}~\bibnamefont {Chen}},\ }\bibfield  {title} {\bibinfo
  {title} {Large nonreciprocal absorption and emission of radiation in type-i
  weyl semimetals with time reversal symmetry breaking},\ }\href
  {https://doi.org/10.1103/PhysRevB.101.165426} {\bibfield  {journal} {\bibinfo
   {journal} {Phys. Rev. B}\ }\textbf {\bibinfo {volume} {101}},\ \bibinfo
  {pages} {165426} (\bibinfo {year} {2020})}\BibitemShut {NoStop}%
\bibitem [{\citenamefont {{Abraham Ekeroth}}\ \emph {et~al.}(2017)\citenamefont
  {{Abraham Ekeroth}}, \citenamefont {Garc{\'{i}}a-Mart{\'{i}}n},\ and\
  \citenamefont {Cuevas}}]{TDDA_2017}%
  \BibitemOpen
  \bibfield  {author} {\bibinfo {author} {\bibfnamefont {R.~M.}\ \bibnamefont
  {{Abraham Ekeroth}}}, \bibinfo {author} {\bibfnamefont {A.}~\bibnamefont
  {Garc{\'{i}}a-Mart{\'{i}}n}},\ and\ \bibinfo {author} {\bibfnamefont {J.~C.}\
  \bibnamefont {Cuevas}},\ }\bibfield  {title} {\bibinfo {title} {{Thermal
  discrete dipole approximation for the description of thermal emission and
  radiative heat transfer of magneto-optical systems}},\ }\href
  {https://doi.org/10.1103/PhysRevB.95.235428} {\bibfield  {journal} {\bibinfo
  {journal} {Phys. Rev. B}\ }\textbf {\bibinfo {volume} {95}},\ \bibinfo
  {pages} {1} (\bibinfo {year} {2017})},\ \Eprint
  {https://arxiv.org/abs/1702.04273} {1702.04273} \BibitemShut {NoStop}%
\bibitem [{\citenamefont {Ashby}\ and\ \citenamefont
  {Carbotte}(2014)}]{Chiral_anomaly-optical_absorption}%
  \BibitemOpen
  \bibfield  {author} {\bibinfo {author} {\bibfnamefont {P.~E.~C.}\
  \bibnamefont {Ashby}}\ and\ \bibinfo {author} {\bibfnamefont {J.~P.}\
  \bibnamefont {Carbotte}},\ }\bibfield  {title} {\bibinfo {title} {Chiral
  anomaly and optical absorption in weyl semimetals},\ }\href
  {https://doi.org/10.1103/PhysRevB.89.245121} {\bibfield  {journal} {\bibinfo
  {journal} {Phys. Rev. B}\ }\textbf {\bibinfo {volume} {89}},\ \bibinfo
  {pages} {245121} (\bibinfo {year} {2014})}\BibitemShut {NoStop}%
\bibitem [{\citenamefont {Hosur}\ and\ \citenamefont
  {Qi}(2015)}]{Tunable-circular-dichroism_WSM}%
  \BibitemOpen
  \bibfield  {author} {\bibinfo {author} {\bibfnamefont {P.}~\bibnamefont
  {Hosur}}\ and\ \bibinfo {author} {\bibfnamefont {X.-L.}\ \bibnamefont {Qi}},\
  }\bibfield  {title} {\bibinfo {title} {Tunable circular dichroism due to the
  chiral anomaly in weyl semimetals},\ }\href
  {https://doi.org/10.1103/PhysRevB.91.081106} {\bibfield  {journal} {\bibinfo
  {journal} {Phys. Rev. B}\ }\textbf {\bibinfo {volume} {91}},\ \bibinfo
  {pages} {081106} (\bibinfo {year} {2015})}\BibitemShut {NoStop}%
\bibitem [{\citenamefont {Resnick}\ \emph {et~al.}(1999)\citenamefont
  {Resnick}, \citenamefont {Persons},\ and\ \citenamefont
  {Lindquist}}]{resnick1999polarized}%
  \BibitemOpen
  \bibfield  {author} {\bibinfo {author} {\bibfnamefont {A.}~\bibnamefont
  {Resnick}}, \bibinfo {author} {\bibfnamefont {C.}~\bibnamefont {Persons}},\
  and\ \bibinfo {author} {\bibfnamefont {G.}~\bibnamefont {Lindquist}},\
  }\bibfield  {title} {\bibinfo {title} {{Polarized emissivity and
  Kirchhoff’s law}},\ }\href@noop {} {\bibfield  {journal} {\bibinfo
  {journal} {Applied optics}\ }\textbf {\bibinfo {volume} {38}},\ \bibinfo
  {pages} {1384} (\bibinfo {year} {1999})}\BibitemShut {NoStop}%
\bibitem [{\citenamefont {Zhang}\ \emph {et~al.}(2020)\citenamefont {Zhang},
  \citenamefont {Wu},\ and\ \citenamefont {Fu}}]{zhang2020validity}%
  \BibitemOpen
  \bibfield  {author} {\bibinfo {author} {\bibfnamefont {Z.~M.}\ \bibnamefont
  {Zhang}}, \bibinfo {author} {\bibfnamefont {X.}~\bibnamefont {Wu}},\ and\
  \bibinfo {author} {\bibfnamefont {C.}~\bibnamefont {Fu}},\ }\bibfield
  {title} {\bibinfo {title} {{Validity of Kirchhoff's law for semitransparent
  films made of anisotropic materials}},\ }\href@noop {} {\bibfield  {journal}
  {\bibinfo  {journal} {Journal of Quantitative Spectroscopy and Radiative
  Transfer}\ }\textbf {\bibinfo {volume} {245}},\ \bibinfo {pages} {106904}
  (\bibinfo {year} {2020})}\BibitemShut {NoStop}%
\bibitem [{\citenamefont {Ott}\ \emph {et~al.}(2018)\citenamefont {Ott},
  \citenamefont {Ben-Abdallah},\ and\ \citenamefont {Biehs}}]{Ott2018}%
  \BibitemOpen
  \bibfield  {author} {\bibinfo {author} {\bibfnamefont {A.}~\bibnamefont
  {Ott}}, \bibinfo {author} {\bibfnamefont {P.}~\bibnamefont {Ben-Abdallah}},\
  and\ \bibinfo {author} {\bibfnamefont {S.-A.}\ \bibnamefont {Biehs}},\
  }\bibfield  {title} {\bibinfo {title} {Circular heat and momentum flux
  radiated by magneto-optical nanoparticles},\ }\href
  {https://doi.org/10.1103/PhysRevB.97.205414} {\bibfield  {journal} {\bibinfo
  {journal} {Phys. Rev. B}\ }\textbf {\bibinfo {volume} {97}},\ \bibinfo
  {pages} {205414} (\bibinfo {year} {2018})}\BibitemShut {NoStop}%
\bibitem [{\citenamefont {Khandekar}\ and\ \citenamefont
  {Jacob}(2019{\natexlab{b}})}]{Chinmay2019}%
  \BibitemOpen
  \bibfield  {author} {\bibinfo {author} {\bibfnamefont {C.}~\bibnamefont
  {Khandekar}}\ and\ \bibinfo {author} {\bibfnamefont {Z.}~\bibnamefont
  {Jacob}},\ }\bibfield  {title} {\bibinfo {title} {Thermal spin photonics in
  the near-field of nonreciprocal media},\ }\href
  {https://doi.org/10.1088/1367-2630/ab494d} {\bibfield  {journal} {\bibinfo
  {journal} {New Journal of Physics}\ }\textbf {\bibinfo {volume} {21}},\
  \bibinfo {pages} {103030} (\bibinfo {year} {2019}{\natexlab{b}})}\BibitemShut
  {NoStop}%
\bibitem [{\citenamefont {Armitage}\ \emph {et~al.}(2018)\citenamefont
  {Armitage}, \citenamefont {Mele},\ and\ \citenamefont
  {Vishwanath}}]{Weyl_RMP}%
  \BibitemOpen
  \bibfield  {author} {\bibinfo {author} {\bibfnamefont {N.~P.}\ \bibnamefont
  {Armitage}}, \bibinfo {author} {\bibfnamefont {E.~J.}\ \bibnamefont {Mele}},\
  and\ \bibinfo {author} {\bibfnamefont {A.}~\bibnamefont {Vishwanath}},\
  }\bibfield  {title} {\bibinfo {title} {Weyl and dirac semimetals in
  three-dimensional solids},\ }\href
  {https://doi.org/10.1103/RevModPhys.90.015001} {\bibfield  {journal}
  {\bibinfo  {journal} {Rev. Mod. Phys.}\ }\textbf {\bibinfo {volume} {90}},\
  \bibinfo {pages} {015001} (\bibinfo {year} {2018})}\BibitemShut {NoStop}%
\bibitem [{\citenamefont {Hofmann}\ and\ \citenamefont
  {Das~Sarma}(2016)}]{Hofmann_2016_SPP}%
  \BibitemOpen
  \bibfield  {author} {\bibinfo {author} {\bibfnamefont {J.}~\bibnamefont
  {Hofmann}}\ and\ \bibinfo {author} {\bibfnamefont {S.}~\bibnamefont
  {Das~Sarma}},\ }\bibfield  {title} {\bibinfo {title} {Surface plasmon
  polaritons in topological weyl semimetals},\ }\href
  {https://doi.org/10.1103/PhysRevB.93.241402} {\bibfield  {journal} {\bibinfo
  {journal} {Phys. Rev. B}\ }\textbf {\bibinfo {volume} {93}},\ \bibinfo
  {pages} {241402} (\bibinfo {year} {2016})}\BibitemShut {NoStop}%
\bibitem [{\citenamefont {Caloz}\ \emph {et~al.}(2018)\citenamefont {Caloz},
  \citenamefont {Al{\`{u}}}, \citenamefont {Tretyakov}, \citenamefont {Sounas},
  \citenamefont {Achouri},\ and\ \citenamefont {Deck-L{\'{e}}ger}}]{Caloz2018}%
  \BibitemOpen
  \bibfield  {author} {\bibinfo {author} {\bibfnamefont {C.}~\bibnamefont
  {Caloz}}, \bibinfo {author} {\bibfnamefont {A.}~\bibnamefont {Al{\`{u}}}},
  \bibinfo {author} {\bibfnamefont {S.}~\bibnamefont {Tretyakov}}, \bibinfo
  {author} {\bibfnamefont {D.}~\bibnamefont {Sounas}}, \bibinfo {author}
  {\bibfnamefont {K.}~\bibnamefont {Achouri}},\ and\ \bibinfo {author}
  {\bibfnamefont {Z.~L.}\ \bibnamefont {Deck-L{\'{e}}ger}},\ }\bibfield
  {title} {\bibinfo {title} {{Electromagnetic Nonreciprocity}},\ }\href
  {https://doi.org/10.1103/PhysRevApplied.10.047001} {\bibfield  {journal}
  {\bibinfo  {journal} {Phys. Rev. Appl.}\ }\textbf {\bibinfo {volume} {10}},\
  \bibinfo {pages} {1} (\bibinfo {year} {2018})}\BibitemShut {NoStop}%
\bibitem [{\citenamefont {Bliokh}\ \emph {et~al.}(2017)\citenamefont {Bliokh},
  \citenamefont {Bekshaev},\ and\ \citenamefont {Nori}}]{Bliokh2017}%
  \BibitemOpen
  \bibfield  {author} {\bibinfo {author} {\bibfnamefont {K.~Y.}\ \bibnamefont
  {Bliokh}}, \bibinfo {author} {\bibfnamefont {A.~Y.}\ \bibnamefont
  {Bekshaev}},\ and\ \bibinfo {author} {\bibfnamefont {F.}~\bibnamefont
  {Nori}},\ }\bibfield  {title} {\bibinfo {title} {Optical momentum, spin, and
  angular momentum in dispersive media},\ }\href
  {https://doi.org/10.1103/PhysRevLett.119.073901} {\bibfield  {journal}
  {\bibinfo  {journal} {Phys. Rev. Lett.}\ }\textbf {\bibinfo {volume} {119}},\
  \bibinfo {pages} {073901} (\bibinfo {year} {2017})}\BibitemShut {NoStop}%
\bibitem [{\citenamefont {Barnett}\ \emph {et~al.}(2016)\citenamefont
  {Barnett}, \citenamefont {Allen}, \citenamefont {Cameron}, \citenamefont
  {Gilson}, \citenamefont {Padgett}, \citenamefont {Speirits},\ and\
  \citenamefont {Yao}}]{Barnett2016}%
  \BibitemOpen
  \bibfield  {author} {\bibinfo {author} {\bibfnamefont {S.~M.}\ \bibnamefont
  {Barnett}}, \bibinfo {author} {\bibfnamefont {L.}~\bibnamefont {Allen}},
  \bibinfo {author} {\bibfnamefont {R.~P.}\ \bibnamefont {Cameron}}, \bibinfo
  {author} {\bibfnamefont {C.~R.}\ \bibnamefont {Gilson}}, \bibinfo {author}
  {\bibfnamefont {M.~J.}\ \bibnamefont {Padgett}}, \bibinfo {author}
  {\bibfnamefont {F.~C.}\ \bibnamefont {Speirits}},\ and\ \bibinfo {author}
  {\bibfnamefont {A.~M.}\ \bibnamefont {Yao}},\ }\bibfield  {title} {\bibinfo
  {title} {On the natures of the spin and orbital parts of optical angular
  momentum},\ }\href {https://doi.org/10.1088/2040-8978/18/6/064004} {\bibfield
   {journal} {\bibinfo  {journal} {Journal of Optics}\ }\textbf {\bibinfo
  {volume} {18}},\ \bibinfo {pages} {064004} (\bibinfo {year}
  {2016})}\BibitemShut {NoStop}%
\bibitem [{\citenamefont {Yang}\ \emph {et~al.}(2020)\citenamefont {Yang},
  \citenamefont {Khosravi},\ and\ \citenamefont {Jacob}}]{yang2020quantum}%
  \BibitemOpen
  \bibfield  {author} {\bibinfo {author} {\bibfnamefont {L.-P.}\ \bibnamefont
  {Yang}}, \bibinfo {author} {\bibfnamefont {F.}~\bibnamefont {Khosravi}},\
  and\ \bibinfo {author} {\bibfnamefont {Z.}~\bibnamefont {Jacob}},\ }\bibfield
   {title} {\bibinfo {title} {Quantum spin operator of the photon},\
  }\href@noop {} {\bibfield  {journal} {\bibinfo  {journal} {arXiv preprint
  arXiv:2004.03771}\ } (\bibinfo {year} {2020})}\BibitemShut {NoStop}%
\bibitem [{\citenamefont {Ranjbar}\ and\ \citenamefont
  {Gill}(2009{\natexlab{b}})}]{Ranjbar2009}%
  \BibitemOpen
  \bibfield  {author} {\bibinfo {author} {\bibfnamefont {B.}~\bibnamefont
  {Ranjbar}}\ and\ \bibinfo {author} {\bibfnamefont {P.}~\bibnamefont {Gill}},\
  }\bibfield  {title} {\bibinfo {title} {{Circular dichroism techniques:
  Biomolecular and nanostructural analyses- A review}},\ }\href
  {https://doi.org/10.1111/j.1747-0285.2009.00847.x} {\bibfield  {journal}
  {\bibinfo  {journal} {Chem. Biol. Drug Des.}\ }\textbf {\bibinfo {volume}
  {74}},\ \bibinfo {pages} {101} (\bibinfo {year}
  {2009}{\natexlab{b}})}\BibitemShut {NoStop}%
\bibitem [{\citenamefont {Gottarelli}\ \emph {et~al.}(2008)\citenamefont
  {Gottarelli}, \citenamefont {Lena}, \citenamefont {Masiero}, \citenamefont
  {Pieraccini},\ and\ \citenamefont {Spada}}]{gottarelli2008use}%
  \BibitemOpen
  \bibfield  {author} {\bibinfo {author} {\bibfnamefont {G.}~\bibnamefont
  {Gottarelli}}, \bibinfo {author} {\bibfnamefont {S.}~\bibnamefont {Lena}},
  \bibinfo {author} {\bibfnamefont {S.}~\bibnamefont {Masiero}}, \bibinfo
  {author} {\bibfnamefont {S.}~\bibnamefont {Pieraccini}},\ and\ \bibinfo
  {author} {\bibfnamefont {G.~P.}\ \bibnamefont {Spada}},\ }\bibfield  {title}
  {\bibinfo {title} {The use of circular dichroism spectroscopy for studying
  the chiral molecular self-assembly: an overview},\ }\href@noop {} {\bibfield
  {journal} {\bibinfo  {journal} {Chirality: The Pharmacological, Biological,
  and Chemical Consequences of Molecular Asymmetry}\ }\textbf {\bibinfo
  {volume} {20}},\ \bibinfo {pages} {471} (\bibinfo {year} {2008})}\BibitemShut
  {NoStop}%
\bibitem [{\citenamefont {Jackson}(1999)}]{jackson1999}%
  \BibitemOpen
  \bibfield  {author} {\bibinfo {author} {\bibfnamefont {J.~D.}\ \bibnamefont
  {Jackson}},\ }\href@noop {} {\emph {\bibinfo {title} {Classical
  Electrodynamics}}},\ \bibinfo {edition} {3rd}\ ed.\ (\bibinfo  {publisher}
  {Wiley},\ \bibinfo {year} {1999})\BibitemShut {NoStop}%
\bibitem [{\citenamefont {Xiao}\ \emph {et~al.}(2020)\citenamefont {Xiao},
  \citenamefont {Wan}, \citenamefont {Shahsafi}, \citenamefont {Salman},
  \citenamefont {Yu}, \citenamefont {Wambold}, \citenamefont {Mei},
  \citenamefont {Perez}, \citenamefont {Derdeyn}, \citenamefont {Yao} \emph
  {et~al.}}]{xiao2020precision}%
  \BibitemOpen
  \bibfield  {author} {\bibinfo {author} {\bibfnamefont {Y.}~\bibnamefont
  {Xiao}}, \bibinfo {author} {\bibfnamefont {C.}~\bibnamefont {Wan}}, \bibinfo
  {author} {\bibfnamefont {A.}~\bibnamefont {Shahsafi}}, \bibinfo {author}
  {\bibfnamefont {J.}~\bibnamefont {Salman}}, \bibinfo {author} {\bibfnamefont
  {Z.}~\bibnamefont {Yu}}, \bibinfo {author} {\bibfnamefont {R.}~\bibnamefont
  {Wambold}}, \bibinfo {author} {\bibfnamefont {H.}~\bibnamefont {Mei}},
  \bibinfo {author} {\bibfnamefont {B.~E.~R.}\ \bibnamefont {Perez}}, \bibinfo
  {author} {\bibfnamefont {W.}~\bibnamefont {Derdeyn}}, \bibinfo {author}
  {\bibfnamefont {C.}~\bibnamefont {Yao}}, \emph {et~al.},\ }\bibfield  {title}
  {\bibinfo {title} {{Precision Measurements of Temperature-Dependent and
  Nonequilibrium Thermal Emitters}},\ }\href@noop {} {\bibfield  {journal}
  {\bibinfo  {journal} {Laser \& Photonics Reviews}\ }\textbf {\bibinfo
  {volume} {14}},\ \bibinfo {pages} {1900443} (\bibinfo {year}
  {2020})}\BibitemShut {NoStop}%
\bibitem [{\citenamefont {Liu}\ \emph {et~al.}(2018)\citenamefont {Liu},
  \citenamefont {Sun}, \citenamefont {Kumar}, \citenamefont {Muechler},
  \citenamefont {Sun}, \citenamefont {Jiao}, \citenamefont {Yang},
  \citenamefont {Liu}, \citenamefont {Liang}, \citenamefont {Xu} \emph
  {et~al.}}]{Co3Sn2S2}%
  \BibitemOpen
  \bibfield  {author} {\bibinfo {author} {\bibfnamefont {E.}~\bibnamefont
  {Liu}}, \bibinfo {author} {\bibfnamefont {Y.}~\bibnamefont {Sun}}, \bibinfo
  {author} {\bibfnamefont {N.}~\bibnamefont {Kumar}}, \bibinfo {author}
  {\bibfnamefont {L.}~\bibnamefont {Muechler}}, \bibinfo {author}
  {\bibfnamefont {A.}~\bibnamefont {Sun}}, \bibinfo {author} {\bibfnamefont
  {L.}~\bibnamefont {Jiao}}, \bibinfo {author} {\bibfnamefont {S.-Y.}\
  \bibnamefont {Yang}}, \bibinfo {author} {\bibfnamefont {D.}~\bibnamefont
  {Liu}}, \bibinfo {author} {\bibfnamefont {A.}~\bibnamefont {Liang}}, \bibinfo
  {author} {\bibfnamefont {Q.}~\bibnamefont {Xu}}, \emph {et~al.},\ }\bibfield
  {title} {\bibinfo {title} {Giant anomalous hall effect in a ferromagnetic
  kagome-lattice semimetal},\ }\href@noop {} {\bibfield  {journal} {\bibinfo
  {journal} {Nature Physics}\ }\textbf {\bibinfo {volume} {14}},\ \bibinfo
  {pages} {1125} (\bibinfo {year} {2018})}\BibitemShut {NoStop}%
\bibitem [{\citenamefont {Xu}\ \emph {et~al.}(2018)\citenamefont {Xu},
  \citenamefont {Liu}, \citenamefont {Shi}, \citenamefont {Muechler},
  \citenamefont {Gayles}, \citenamefont {Felser},\ and\ \citenamefont
  {Sun}}]{Co3Sn2S2_Co3Sn2Se2}%
  \BibitemOpen
  \bibfield  {author} {\bibinfo {author} {\bibfnamefont {Q.}~\bibnamefont
  {Xu}}, \bibinfo {author} {\bibfnamefont {E.}~\bibnamefont {Liu}}, \bibinfo
  {author} {\bibfnamefont {W.}~\bibnamefont {Shi}}, \bibinfo {author}
  {\bibfnamefont {L.}~\bibnamefont {Muechler}}, \bibinfo {author}
  {\bibfnamefont {J.}~\bibnamefont {Gayles}}, \bibinfo {author} {\bibfnamefont
  {C.}~\bibnamefont {Felser}},\ and\ \bibinfo {author} {\bibfnamefont
  {Y.}~\bibnamefont {Sun}},\ }\bibfield  {title} {\bibinfo {title} {Topological
  surface fermi arcs in the magnetic weyl semimetal
  {${\mathrm{Co}}_{3}{\mathrm{Sn}}_{2}{\mathrm{S}}_{2}$}},\ }\href
  {https://doi.org/10.1103/PhysRevB.97.235416} {\bibfield  {journal} {\bibinfo
  {journal} {Phys. Rev. B}\ }\textbf {\bibinfo {volume} {97}},\ \bibinfo
  {pages} {235416} (\bibinfo {year} {2018})}\BibitemShut {NoStop}%
\bibitem [{\citenamefont {Greffet}\ \emph {et~al.}(2018)\citenamefont
  {Greffet}, \citenamefont {Bouchon}, \citenamefont {Brucoli},\ and\
  \citenamefont {Marquier}}]{Greffet2018}%
  \BibitemOpen
  \bibfield  {author} {\bibinfo {author} {\bibfnamefont {J.~J.}\ \bibnamefont
  {Greffet}}, \bibinfo {author} {\bibfnamefont {P.}~\bibnamefont {Bouchon}},
  \bibinfo {author} {\bibfnamefont {G.}~\bibnamefont {Brucoli}},\ and\ \bibinfo
  {author} {\bibfnamefont {F.}~\bibnamefont {Marquier}},\ }\bibfield  {title}
  {\bibinfo {title} {{Light Emission by Nonequilibrium Bodies: Local Kirchhoff
  Law}},\ }\href {https://doi.org/10.1103/PhysRevX.8.021008} {\bibfield
  {journal} {\bibinfo  {journal} {Phys. Rev. X}\ }\textbf {\bibinfo {volume}
  {8}},\ \bibinfo {pages} {21008} (\bibinfo {year} {2018})}\BibitemShut
  {NoStop}%
\bibitem [{\citenamefont {Edalatpour}\ and\ \citenamefont
  {Francoeur}(2014)}]{Edalatpour2014}%
  \BibitemOpen
  \bibfield  {author} {\bibinfo {author} {\bibfnamefont {S.}~\bibnamefont
  {Edalatpour}}\ and\ \bibinfo {author} {\bibfnamefont {M.}~\bibnamefont
  {Francoeur}},\ }\bibfield  {title} {\bibinfo {title} {The thermal discrete
  dipole approximation (t-dda) for near-field radiative heat transfer
  simulations in three-dimensional arbitrary geometries},\ }\href
  {https://doi.org/https://doi.org/10.1016/j.jqsrt.2013.08.021} {\bibfield
  {journal} {\bibinfo  {journal} {Journal of Quantitative Spectroscopy and
  Radiative Transfer}\ }\textbf {\bibinfo {volume} {133}},\ \bibinfo {pages}
  {364} (\bibinfo {year} {2014})}\BibitemShut {NoStop}%
\bibitem [{\citenamefont {Edalatpour}\ \emph {et~al.}(2015)\citenamefont
  {Edalatpour}, \citenamefont {\ifmmode~\check{C}\else \v{C}\fi{}uma},
  \citenamefont {Trueax}, \citenamefont {Backman},\ and\ \citenamefont
  {Francoeur}}]{Edalatpour2015}%
  \BibitemOpen
  \bibfield  {author} {\bibinfo {author} {\bibfnamefont {S.}~\bibnamefont
  {Edalatpour}}, \bibinfo {author} {\bibfnamefont {M.}~\bibnamefont
  {\ifmmode~\check{C}\else \v{C}\fi{}uma}}, \bibinfo {author} {\bibfnamefont
  {T.}~\bibnamefont {Trueax}}, \bibinfo {author} {\bibfnamefont
  {R.}~\bibnamefont {Backman}},\ and\ \bibinfo {author} {\bibfnamefont
  {M.}~\bibnamefont {Francoeur}},\ }\bibfield  {title} {\bibinfo {title}
  {Convergence analysis of the thermal discrete dipole approximation},\ }\href
  {https://doi.org/10.1103/PhysRevE.91.063307} {\bibfield  {journal} {\bibinfo
  {journal} {Phys. Rev. E}\ }\textbf {\bibinfo {volume} {91}},\ \bibinfo
  {pages} {063307} (\bibinfo {year} {2015})}\BibitemShut {NoStop}%
\bibitem [{\citenamefont {Gao}\ \emph {et~al.}(2021)\citenamefont {Gao},
  \citenamefont {Khandekar}, \citenamefont {Jacob},\ and\ \citenamefont
  {Li}}]{Gao2020}%
  \BibitemOpen
  \bibfield  {author} {\bibinfo {author} {\bibfnamefont {X.}~\bibnamefont
  {Gao}}, \bibinfo {author} {\bibfnamefont {C.}~\bibnamefont {Khandekar}},
  \bibinfo {author} {\bibfnamefont {Z.}~\bibnamefont {Jacob}},\ and\ \bibinfo
  {author} {\bibfnamefont {T.}~\bibnamefont {Li}},\ }\bibfield  {title}
  {\bibinfo {title} {Thermal equilibrium spin torque: Near-field radiative
  angular momentum transfer in magneto-optical media},\ }\href
  {https://doi.org/10.1103/PhysRevB.103.125424} {\bibfield  {journal} {\bibinfo
   {journal} {Phys. Rev. B}\ }\textbf {\bibinfo {volume} {103}},\ \bibinfo
  {pages} {125424} (\bibinfo {year} {2021})}\BibitemShut {NoStop}%
\bibitem [{\citenamefont {Messina}\ \emph {et~al.}(2013)\citenamefont
  {Messina}, \citenamefont {Tschikin}, \citenamefont {Biehs},\ and\
  \citenamefont {Ben-Abdallah}}]{multiple_dipoles_TDDA_2013}%
  \BibitemOpen
  \bibfield  {author} {\bibinfo {author} {\bibfnamefont {R.}~\bibnamefont
  {Messina}}, \bibinfo {author} {\bibfnamefont {M.}~\bibnamefont {Tschikin}},
  \bibinfo {author} {\bibfnamefont {S.-A.}\ \bibnamefont {Biehs}},\ and\
  \bibinfo {author} {\bibfnamefont {P.}~\bibnamefont {Ben-Abdallah}},\
  }\bibfield  {title} {\bibinfo {title} {Fluctuation-electrodynamic theory and
  dynamics of heat transfer in systems of multiple dipoles},\ }\href
  {https://doi.org/10.1103/PhysRevB.88.104307} {\bibfield  {journal} {\bibinfo
  {journal} {Phys. Rev. B}\ }\textbf {\bibinfo {volume} {88}},\ \bibinfo
  {pages} {104307} (\bibinfo {year} {2013})}\BibitemShut {NoStop}%
\end{thebibliography}%






\end{document}